\documentclass[twocolumn]{aastex631}
\usepackage{amsmath}
\usepackage{enumitem}
\usepackage{xspace}

\usepackage{float}

\newcommand{\M}[0]{\ensuremath{\rm M_{\rm 1.4}}\xspace}

\setlist{
  noitemsep,
  listparindent=\parindent,
  parsep=0pt,
}
\usepackage[at]{easylist}
\usepackage{multirow}
\begin{document}

\title{The first systematically identified repeating partial tidal disruption event}

\author[0000-0001-8426-5732]{Jean J. Somalwar}
\affil{Cahill Center for Astronomy and Astrophysics, MC\,249-17 California Institute of Technology, Pasadena CA 91125, USA.}

\author[0000-0002-7252-5485]{Vikram Ravi}
\affil{Cahill Center for Astronomy and Astrophysics, MC\,249-17 California Institute of Technology, Pasadena CA 91125, USA.}

\author[0000-0001-6747-8509]{Yuhan Yao}
\affiliation{Miller Institute for Basic Research in Science, 468 Donner Lab, Berkeley, CA 94720, USA}
\affiliation{Department of Astronomy, University of California, Berkeley, CA 94720, USA}

\author[0000-0002-5063-0751]{Muryel Guolo}
\affiliation{Bloomberg Center for Physics and Astronomy, Johns Hopkins University, 3400 N. Charles St., Baltimore, MD 21218, USA}

\author{Matthew Graham}
\affil{Cahill Center for Astronomy and Astrophysics, MC\,249-17 California Institute of Technology, Pasadena CA 91125, USA.}

\author[0000-0002-5698-8703]{Erica Hammerstein}
\affiliation{Department of Astronomy, University of Maryland, College Park, MD 20742, USA}

\author[0000-0002-1568-7461]{Wenbin Lu}
\affiliation{Department of Astronomy and Theoretical Astrophysics Center, University of California, Berkeley, CA 94720, USA}

\author[0000-0002-2555-3192]{Matt Nicholl}
\affiliation{Astrophysics Research Centre, School of Mathematics and Physics, Queens University Belfast, Belfast BT7 1NN, UK}

\author[0000-0003-4531-1745]{Yashvi Sharma}
\affiliation{Cahill Center for Astronomy and Astrophysics, MC\,249-17 California Institute of Technology, Pasadena CA 91125, USA.}

\author{Robert Stein}
\affil{Cahill Center for Astronomy and Astrophysics, MC\,249-17 California Institute of Technology, Pasadena CA 91125, USA.}

\author[0000-0002-3859-8074]{Sjoert van Velzen}
\affiliation{Leiden Observatory, Leiden University, Postbus 9513, 2300 RA, Leiden, The Netherlands}

\author[0000-0001-8018-5348]{Eric C. Bellm} 
\affiliation{DIRAC Institute, Department of Astronomy, University of Washington, 3910 15th Avenue NE, Seattle, WA 98195, USA}

\author[0000-0002-8262-2924]{Michael W. Coughlin}
\affiliation{School of Physics and Astronomy, University of Minnesota, Minneapolis, Minnesota 55455, USA}

\author[0000-0001-5668-3507]{Steven L. Groom}
\affiliation{IPAC, California Institute of Technology, 1200 E. California
             Blvd, Pasadena, CA 91125, USA}

\author[0000-0002-8532-9395]{Frank J. Masci}
\affiliation{IPAC, California Institute of Technology, 1200 E. California
             Blvd, Pasadena, CA 91125, USA}

\author[0000-0002-0387-370X]{Reed Riddle}
\affiliation{Caltech Optical Observatories, California Institute of Technology, Pasadena, CA 91125, USA}




\begin{abstract}
Tidal disruption events (TDEs) occur when a star enters the tidal radius of a supermassive black hole (SMBH). If the star only grazes the tidal radius, a fraction of the stellar mass will be accreted in a partial TDE (pTDE). The remainder can continue orbiting and may re-disrupted at pericenter, causing a repeating pTDE. pTDEs may be as or more common than full TDEs (fTDEs), yet few are known. In this work, we present the discovery of the first repeating pTDE from a systematically-selected sample, AT\,2020vdq. AT\,2020vdq was originally identified as an optically- and radio-flaring TDE. Around $3$ years after its discovery, it rebrightened dramatically and rapidly in the optical. The optical flare was remarkably fast and luminous compared to previous TDEs. It was accompanied by extremely broad (${\sim}0.1c$) optical/UV spectral features and faint X-ray emission ($L_X \sim 3\times10^{41}$\,erg\,s$^{-1}$), but no new radio-emitting component. Based on the transient optical/UV spectral features and the broadband light curve, we show that AT\,2020vdq is a repeating pTDE. We then use it to constrain TDE models; in particular, we favor a star originally in a very tight binary system that is tidally broken apart by the Hills mechanism. We also constrain the repeating pTDE rate to be $10^{-6}$ to $10^{-5}$ yr$^{-1}$ galaxy$^{-1}$, with uncertainties dominated by the unknown distribution of pTDE repeat timescales. In the Hills framework, this means the binary fraction in the galactic nucleus is of the order few percent.
\end{abstract}

\section{Introduction} \label{sec:intro}

Much about supermassive black holes (SMBHs) remains enigmatic, including their formation pathways, typical growth histories, and effects on their hosts \citep[e.g.][]{kormendoho_review}. Much of the reason for our lack of knowledge about SMBHs is their inherent faintness: if the SMBH is not actively accreting as an active galactic nucleus (AGN), it can only be detected if it is in a very nearby galaxy. Even if it is accreting, the accretion process is poorly understood, so inferring physical properties of the SMBH and its environment from the observed emission is nontrivial. 

In the last few decades, tidal disruption events (TDEs) have become key probes of SMBH physics \citep[see][for a review]{suvi_review}. TDEs occur when a star ventures within the tidal radius of an SMBH: $R_T \approx R_* (M_{\rm BH}/M_*)^{1/3}$ , where $R_T$ is the tidal radius, $M_{\rm BH}$ is the SMBH mass, $R_*$ is the stellar radius, and $M_*$ is the stellar mass \citep[][]{Evans1989ApJ...346L..13E, Rees1988Natur.333..523R, Phinney1989IAUS..136..543P}. Tidal forces shred the star and, eventually, the stellar debris is accreted, producing a bright multiwavelength flare. The development of high cadence, wide field optical, radio, and X-ray surveys has enabled the discovery of ${\gtrsim}100$ candidate events \citep[e.g.][]{erosita_xray, Yao_tdes}.

To produce a TDE, a star must enter into a plunging orbit with a pericenter that is within the tidal radius of the SMBH but outside the Schwarzschild radius. This occurs only once every ${\sim}10^{4-5}$ years in galaxies with $M_{\rm BH} \lesssim 10^8\,M_\odot$ \citep[][]{sjoert_tderate,2020SSRv..216...35S,Yao_tdes}. In contrast, if the star is on a grazing orbit, but still reaches the vicinity of the tidal radius, it can be partially disrupted, and only a fraction of the stellar mass will be deposited on the SMBH \citep[][]{guillochon_ramirez_ptdehydro,coughlin_nixon_ptde}. These partial TDEs (pTDEs) may occur only once if the star is on a parabolic orbit, or they can repeat if the star is on an elliptical orbit. Theoretical work has suggested that the pTDE rate could be orders of magnitude higher than the full TDE (fTDE) rate and may contribute significantly to the growth of SMBHs \citep[][]{Bortolas_ptderates}. They also provide a unique avenue to test/improve our TDE theory: the only things that may change between disruptions are the structure of the star (as it is tidally perturbed during each partial disruption), the amount of mass joining the accretion flow, and the circumnuclear medium (CNM; if the previous disruptions launched outflows). Any differences between subsequent pTDE flares can thus be used to rule out models that invoke any of the static aspects of the system. Moreover, an accretion disk may form after the first disruptions, so subsequent disruptions can be used to assess the role that the disk plays in the TDE behavior.

Few candidate pTDEs are confidently identified, despite the high rate estimates. Given our insufficient understanding of TDE physics, one of the best ways to confirm a candidate pTDE is if it repeats; i.e. if the star is on a bound orbit.
If the orbital period is {$P\simeq 10^5\mathrm{\,yr}\, (a/\mathrm{pc})^{3/2} (M/10^6M_\odot)^{-1/2}$ ($a$ being the orbital semi-major axis and $M$ being the SMBH mass)}, as expected for isolated stars in a nuclear star cluster slowly migrating onto grazing orbits due to orbital relaxation, we will not detect a repeat. It is possible that some or most of the known TDEs are this type of pTDE, but we require a better understanding of pTDE and fTDE physics to be able to accurately classify such sources.

Indeed, the pTDE candidates that are known were discovered because they repeated within ${\sim}0.3-3$ years (ASASSN 14ko \citealp[][]{14ko_discovery,j0456_ptde,fyk_ptde}; eRASSt J045650.3-203750 \citealp{j0456_ptde}; AT 2018fyk \citealp{fyk_discovery,fyk_ptde}; RX J133157.6-324319.7 \citealp{J1331_discovery, J1331_ptde}). Such tightly bound stars are difficult to {reproduce with orbital relaxation alone \citep{alexander17_stellar_dynamics}}. The most likely mechanism for such objects invokes Hills break-up of a tight binary, resulting in one bound object and one ejected object. If the binary is tight enough (inner semi-major axis ${\lesssim}0.01$ AU, with dependence on the SMBH mass), dynamical arguments predict that the bound star will have a {period from $\mathcal{O}(10^2)$\,days to a few years \citep{lu23_QPE_pTDEs}}. The binary fraction in galactic nuclei is not well known, but could be sufficiently high to explain the rate of these events.

These pTDE candidates are an inhomogeneous group, however, with some selected in the X-ray, some in the optical, and none as part of a uniform search. Hence, the pTDE rate and the differences in their multiwavelength properties relative to fTDEs are still unconstrained. Fortunately, over the last few years, the Zwicky Transient Facility (ZTF; \citealp{ztf1, ztf2, ztf3, ztf4}) has enabled systematic searches for optically-flaring TDEs. It has produced the first uniformly-selected samples of TDEs \citep[][]{2021ApJ...908....4V,Hammerstein_tdes,Yao_tdes}. In May 2023, one of the ZTF TDEs, AT\,2020vdq, rebrightened \citep[][]{2023TNSAN.115....1C}. We commenced follow-up efforts and classified it as a repeating pTDE.

In this work, we will present observations and analysis of AT\,2020vdq. We describe the detection and properties of the first flare from AT\,2020vdq in Section~\ref{sec:vdq_initial} and we detail the second flare in Section~\ref{sec:vdq_rebright}. We constrain past flares from this source in Section~\ref{sec:prev_flares}. In Section~\ref{sec:discussion}, we summarize key results, show that AT\,2020vdq is consistent with being a pTDE, compare it to published observations of pTDEs, and use our observations to constrain both TDE and pTDE models. In particular, we constrain key aspects of TDE radio, optical, and spectral line emission, and we constrain the pTDE rate and typical number of repetitions. We also discuss possible contributions of pTDEs to solving the missing energy problem. Finally, we conclude in Section~\ref{sec:conc}.

\begin{deluxetable}{c | cc}
\centerwidetable
\tablewidth{10pt}
\tablecaption{ \label{tab:host}}
\tablehead{ \colhead{Parameter} & \colhead{AT 2020vqd}}
\startdata
R.A. & $10^{\rm h}08^{\rm m}53.50^{\rm s}$ \\
Dec. & $+42^{\rm d}43^{\rm m}00.40^{\rm s}$\\
Redshift & 0.045 \\
$d_L$ [Mpc] & 206
\enddata
\tablecomments{Basic properties of the host galaxy of AT\,2020vdq.}
\centering
\end{deluxetable}

\section{Observation of the initial flare} \label{sec:vdq_initial}

\begin{figure}
    \centering
    \includegraphics[width=0.5\textwidth]{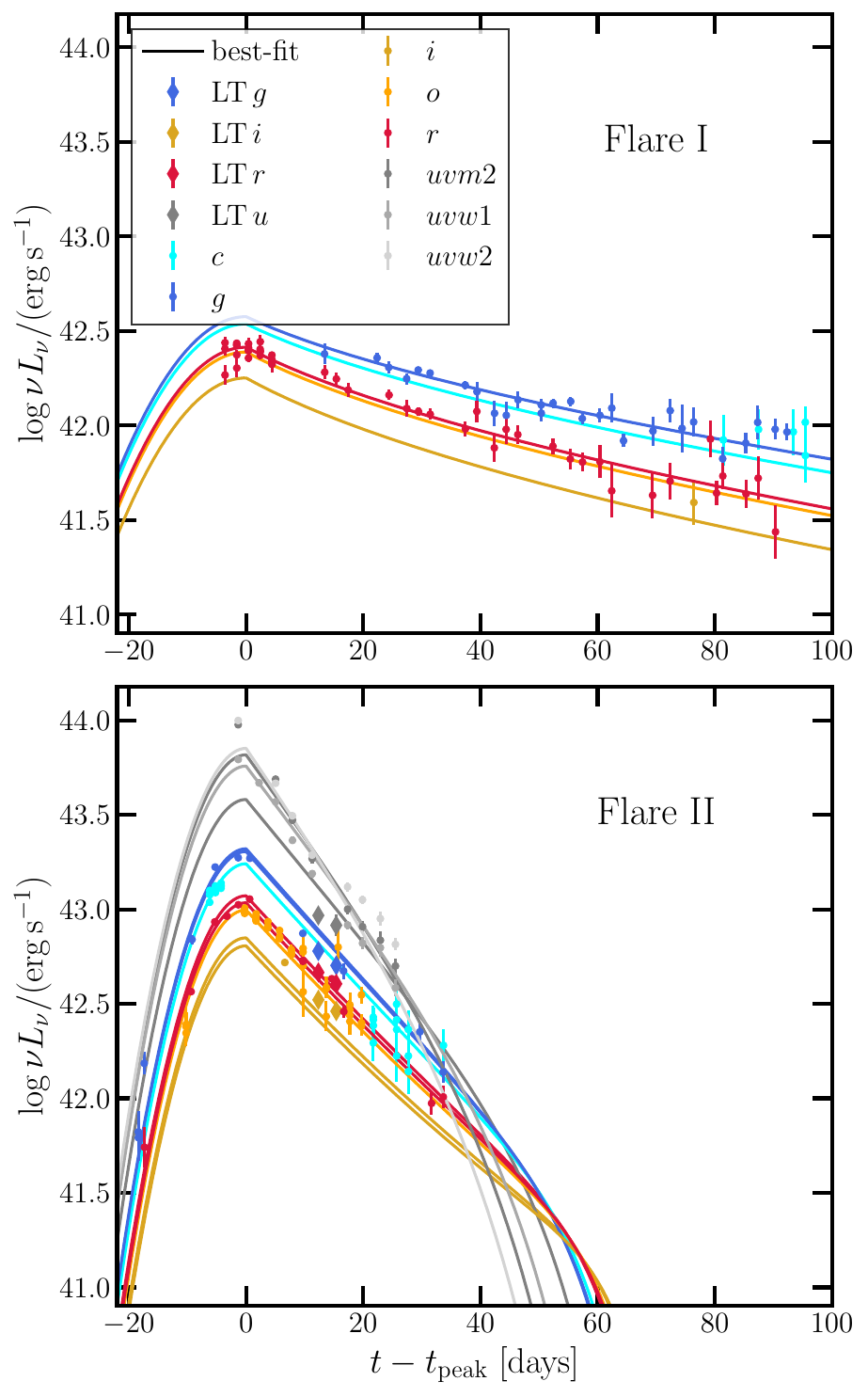}
    \caption{The optical and UV lightcurves for the first ({\it top}) and second ({\it bottom}) flares from AT\,2020vdq. Observations are shown as scattered points. The best-fit parameteric, evolving black body models, as described in Section~\ref{sec:optcon} are shown as solid lines. }
    \label{fig:opt_lc}
\end{figure}

We begin by describing the multiwavelength properties of the first flare from AT\,2020vdq. These properties are described in detail by \cite{paperI, paperII, Yao_tdes}.

AT\,2020vdq was first detected as an optical transient by the Zwicky Transient Facility (ZTF) on Oct. 4 2020 (MJD 59126) using the selection described in \cite{Yao_tdes} implemented using the AMPEL filter \citep[][]{2019A&A...631A.147N}. It was located at the nucleus of a dwarf galaxy at $z = 0.045$ with $\log M_*/M_\odot = 9.25$. The stellar velocity dispersion of the nucleus host galaxy was measured to be $\sigma_* = 44 \pm 3$ km s$^{-1}$ from a high resolution spectrum obtained with ESI on the Keck II telescope \citep[][]{Yao_tdes}. This dispersion corresponds to a black hole mass of $\log M_{\rm BH, \sigma_*}/M_\odot = 5.59 \pm 0.37$ using the $M_{\rm BH}-\sigma_*$ relation from \cite{kormendy_ho}. We refer the reader to \cite{Yao_tdes,paperI,paperII} for more details of the host observations and analysis.

The first optical flare from AT2020vdq is shown in the top panel of Figure~\ref{fig:opt_lc}. The optical flare peaked in the $r$-band near Sept. 23 2020 (MJD 59115). This peak was fit to a blackbody by \cite{Yao_tdes}, who found that the peak is consistent with a blackbody with temperature $\log T_{\rm bb}/{\rm K} = 4.16$ and luminosity $\log L_{\rm bb}/({\rm erg\,s}^{-1}) = 42.99$. Note that this measurement is based on observations in only the $g$ and $r$ bands, so unknown systematic errors render the results uncertain. Stronger constraints are not possible because no UV observations were obtained at peak. This luminosity makes AT\,2020vdq the lowest luminosity TDE in the \cite{Yao_tdes} sample, even relative to events with lower black hole masses. The flare temperature is also in the coolest ~20\% quantile of the sample. 

\begin{figure}
    \centering
    \includegraphics[width=0.5\textwidth]{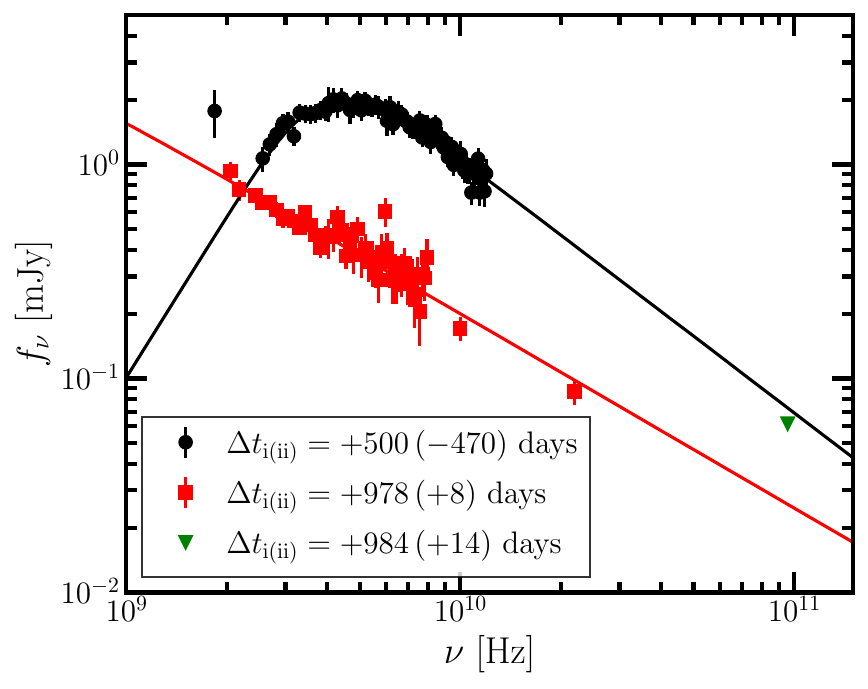}
    \caption{The radio evolution of AT\,2020vdq. The black scatter points show a VLA observation from 500 days after the initial flare, or equivalently 470 days before the second flare. The red scatter points show a VLA observations from eight days after the rebrightening. The green upper limit shows a NOEMA observation from two weeks after the rebrightening. Both SEDs can be fit with spherical, non-relativistic synchrotron models. No young emitting component is required in the SED from shortly after the rebrightening.}
    \label{fig:radio_sed}
\end{figure}

AT\,2020vdq was detected as a radio source in 3 GHz observations from the VLA Sky Survey (VLASS; \citealp{Lacy}) on Oct. 9 2021 (MJD 59496), or ${\sim}1$ year after the first optical peak. It had a flux of $f_{\nu, {\rm 3\,GHz}} = 1.48\pm0.14\,$mJy or $\nu L_{\nu} = (2.3\pm0.2)\times 10^{38}\,{\rm erg}\,{\rm s}^{-1}$. The transient location had been previously observed by VLASS on May 25, 2019 (MJD 58628), or ${\sim}1.3$ years before the optical peak. No emission was detected, with a $3\sigma$ upper limit $f_{\nu, {\rm 3\,GHz}} < 0.7\,$mJy or $\nu L_{\nu} < 10^{38}\,{\rm erg}\,{\rm s}^{-1}$. We obtained follow-up radio observations with the VLA on Feb. 2, 2022 (MJD 59612), or $1.4$ years after the optical peak. The radio SED, shown in Figure~\ref{fig:radio_sed} was consistent with a non-relativistic, spherical synchrotron source with equipartition radius $\log R_{\rm eq}/{\rm cm} = 17.07^{+0.01}_{-0.01}$ and magnetic field $\log B_{\rm eq}/{\rm G} = -0.59^{+0.05}_{-0.04}$. Assuming the radio-emitting outflow was launched near optical peak, this radius corresponds to an average velocity $v\approx2.7\times10^4\,$km\,s$^{-1}$, or $\beta=v/c\approx0.09$. The energy in the outflow is $\log E_{\rm eq}/{\rm erg} = 49.7^{+0.01}_{-0.01}$. The power-law index of the electron energy distribution was consistent with $p = 3.4^{+0.1}_{-0.1}$. We refer the reader to \cite{paperI,paperII} for more details of the radio modeling and observations.

AT\,2020vdq was not detected as an X-ray transient in observations with the Swift/XRT telescope on Feb. 28, 2022; the $3\sigma$ upper limit on the $0.2-10$ keV flux was $L_X \lesssim 3\times 10^{42}$ erg s$^{-1}$. No significant transient UV emission was detected by Swift/UVOT on the same date.

We obtained optical spectra of AT\,2020vdq on Feb. 6 and Apr. 7, 2022 (MJDs 59616, 59676), or 1.4/1.5 years post-optical peak using the LRIS spectrograph on the Keck I telescope (for reference, the source repeated after $2.7$ years). The observation and reduction details are described in \cite{paperI,paperII}. The spectra are shown and discussed in detail in \cite{paperII}. Multiple transient spectral features are visible, as described in detail in \cite{paperII}: intermediate width Balmer, He\,II$\lambda 4685$, He\,I$\lambda 5875$, and [Fe\,X]$\lambda6374$ features. The Balmer features have asymmetric profiles with a slightly redshifted peak and a blue tail with average FWHM${\sim}900$\,km\,s$^{-1}$. The Balmer emission is consistent with being produced by dense $n_e \gtrsim 10^5$ cm$^{-3}$, compact $R\lesssim 10^{17}$ cm gas. The Balmer decrement of these lines is high: H${\rm \alpha}$/H${\rm \beta} = 9$, suggesting that the emission is heavily extincted or produced by a non-standard ionization mechanism. The Balmer features brightened between the two spectra. The other transient features have narrower, symmetric profiles with FWHM${\sim}300$ km s$^{-1}$.

\section{Observations of the rebrightening} \label{sec:vdq_rebright}

\subsection{Broadband transient optical/UV emission} \label{sec:optcon}

\begin{deluxetable*}{c|ccccccccc}
\centerwidetable
\tablewidth{10pt}
\tablecaption{ \label{tab:optlc}}
\tablehead{ \colhead{} & \colhead{$t_{\rm peak}$} & \colhead{$\log \frac{L_{\rm bb}}{{\rm erg\,s}^{-1}}$} & \colhead{$\lambda_{\rm Edd.}$} & \colhead{$\log T_{\rm bb}/{\rm K}$} & \colhead{$\frac{dT}{dt}$} & \colhead{$t_{\rm 1/2,rise}$} & \colhead{$t_{\rm 1/2,decay}$} & \colhead{$\log \frac{E_{\rm bb}}{{\rm erg}}$} &  \colhead{$\log \frac{M_{\rm bb}}{{\rm M_\odot}}$} \cr 
 \colhead{} & \colhead{} & \colhead{} & \colhead{} & \colhead{} & \colhead{[K/day]} & \colhead{[day]} & \colhead{[day]} & \colhead{} &  \colhead{}}
\startdata
Flare I & $59113.1 \pm 0.8$ & $42.78 \pm 0.06$ & $0.12 \pm 0.02$ & $4.02 \pm 0.04$ & $66 \pm 19$ & $7.8 \pm 5.6$ & $48 \pm 16$ & $49.8$ & $-4.1$ \cr
Flare II & $60082.60 \pm 0.07$ & $44.00 \pm 0.01$ & $1.99 \pm 0.05$ & $4.317 \pm 0.004$ & $-249 \pm 16$ & $4.67 \pm 0.06$ & $6.4 \pm 0.1$ & $51.3$ & $-2.7$
\enddata
\tablecomments{Best-fit evolving blackbody parameters for the optical flares from AT\,2020vdq.}
\centering
\end{deluxetable*}

The rebrightening of AT\,2020vdq was first detected by optical surveys: ZTF detected the rebrightening on May 9, 2023 (MJD 60073), 2.6 years after the first optical peak. This optical flare is shown in the bottom panel of Figure~\ref{fig:opt_lc}. UV observations with Swift/UVOT were triggered shortly after (PIs: Lin, Leloudas, Guolo). We processed this data using the same methods as \cite{sjoert_tdes,Hammerstein_tdes}; i.e., we reduced the Swift/UVOT data using recommended procedures, measured the source flux using aperture photometry, and subtracted the host contribution using the results from spectral energy distribution fits to the host galaxy from \cite{paperI}. The UV lightcurve is overlaid on Figure~\ref{fig:opt_lc}. Optical ($ugri$) observations with the Liverpool telescope observations were also triggered (PI: Nicholl). The observations were reduced using standard methods and the host component was subtracted using SDSS photometry.
The resulting difference photometry is overlaid on Figure~\ref{fig:opt_lc}. The evolution of this optical/UV rebrightening is notably different from that of the first flare. The emission peaks at a $g$-band luminosity a factor ${\sim}5$ higher than that of the original flare. Both the rise and decay times of the rebrightening are significantly faster than those of the original flare. The rebrightening also cools significantly within ${\sim}10$ days post-peak. Similar cooling is not obviously visible during the first flare. 

We quantify these differences by modelling the optical/UV emission with two methods. First, we fit the UV/optical lightcurves with a parametric model, following \cite{Yao_tdes}. We assume the lightcurve can be modeled as an evolving blackbody. The blackbody luminosity rises as a Gaussian and decays as a power law and the temperature is constant pre-peak and evolves linearly post-peak:

\begin{gather}
    L_{\rm bb} = L_{\rm peak} \begin{cases}
        e^{-\frac{1}{2}\big(\frac{t-t_{\rm peak}}{\sigma_{\rm rise}}\big)^2}, & t<t_{\rm peak}\\
        \big(\frac{t-t_{\rm peak+t_0}}{t_0}\big)^p, & \text{otherwise.}
    \end{cases}\\
    T_{\rm bb} = T_0 + \begin{cases}
        0, & t<t_{\rm peak}\\
        \frac{dT}{dt}\times (t-t_{\rm peak}), & \text{otherwise.}
    \end{cases}
\end{gather}
The time of luminosity peak is given by $t_{\rm peak}$, $t_0$ and $p$ control the luminosity decay shape, $\sigma_{\rm rise}$ controls the rate of the rise, and $\frac{dT}{dt}$ controls the rate at which the temperature evolves. We fit both the initial and second flares using this model so that we can compare the temperature evolution. The resulting best-fit parameters are shown in Table~\ref{tab:optlc}. Rather than report $t_0$, $p$, and $\sigma_{\rm rise}$, we show the rise time from half-peak to peak luminosity, $t_{\rm 1/2, rise}$, and the corresponding value for the decay ($t_{\rm 1/2, decay}$). The peak luminosity of the rebrightening is $1.2$ dex brighter than that of the initial flare. The optical/UV Eddington ratio $\lambda = L_{\rm peak}/L_{\rm Edd}$ is correspondingly higher, such that the rebrightening corresponds to super-Eddington emission. The blackbody temperature of the rebrightening is $0.3$ dex higher than that of the first flare, although we caution that the lack of UV observations of the first flare render this measurement uncertain. The rebrightening shows significant cooling, whereas the first flare shows little temperature evolution. Both the rise and fade times of the rebrightening are also significantly faster than those of the first flare.

\begin{figure*}
    \centering
    \includegraphics[width=\textwidth]{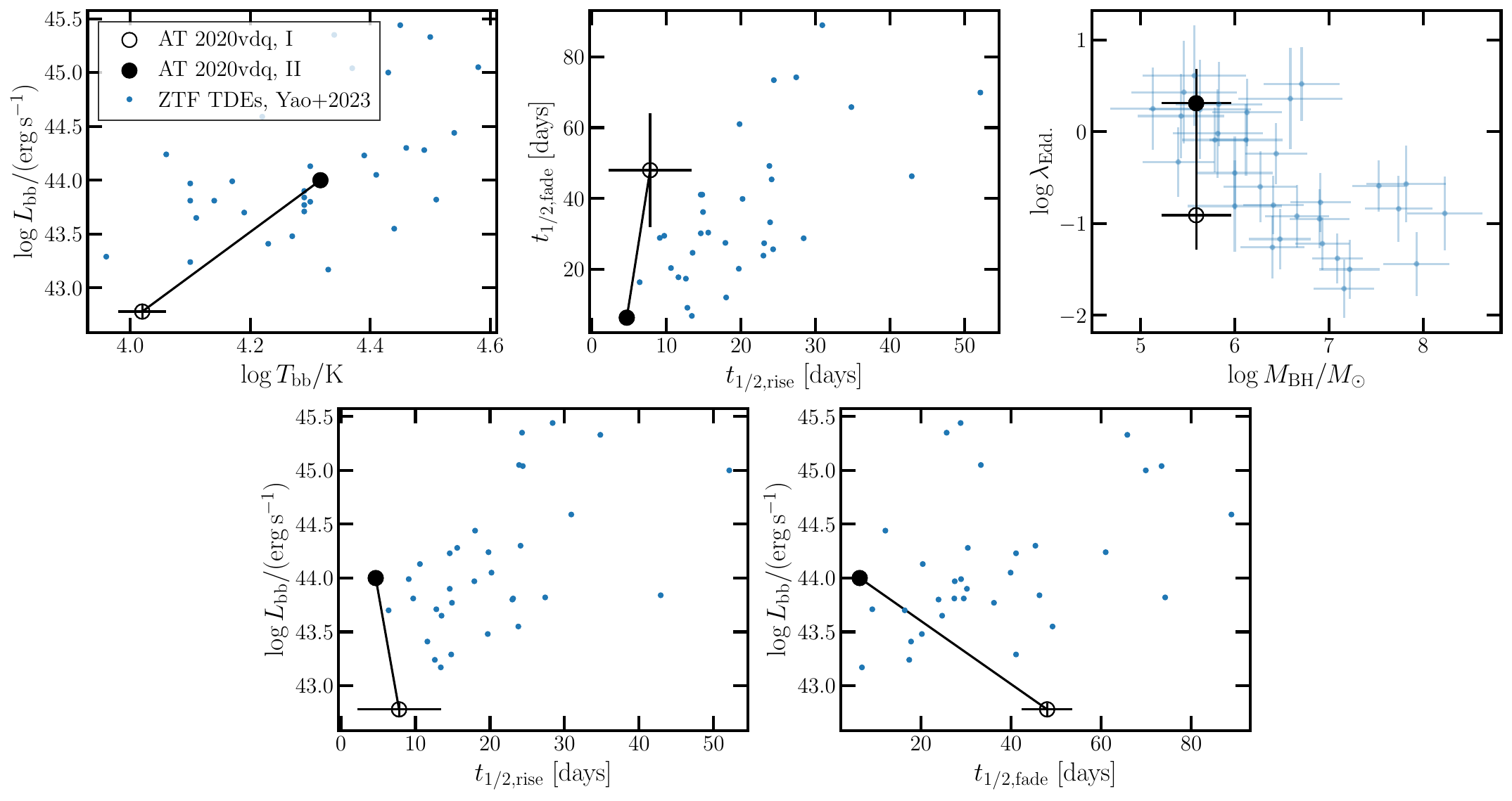}
    \caption{A comparison of the parametric, evolving, blackbody fit parameters for AT\,2020vdq (black) to those of the TDEs from \cite{Yao_tdes} (blue). The initial flare from AT\,2020vdq is denoted with open markers while the rebrightening is denoted with filled markers. The pTDE has a lightcurve that is generally consistent with the broader TDE population, although the rebrightening is luminous and fast evolving. }
    \label{fig:TDE_comp}
\end{figure*}

We compare these best-fit parameters to the ZTF-selected TDEs from \cite{Yao_tdes} in Figure~\ref{fig:TDE_comp}. In the top left panel, we show the evolution of the pTDE in blackbody luminosity/temperature space. The initial flare is remarkably faint and cool, whereas the rebrightening is relatively typical. In the top middle panel, we show the evolution of the rise/decay times. Both the initial and second flares have short rise times relative to the ZTF TDE sample, although the uncertainties on the rise time of the initial flare are large. The second flare, in contrast, is the fastest rising TDE observed by ZTF and it has one of the fastest decay times. In the top right panel, we show the evolution of the flare Eddington ratio. The peak Eddington ratio of the first flare is significantly lower than that of the ZTF TDEs at similar black hole masses. The second flare, on the other hand, is at the upper end of the Eddington ratio distribution. In the bottom left panel, we show the rise time versus peak luminosity. In this parameter space, it is clear that AT\,2020vdq is both fast-rising and the rebrightening is very luminous for this fast rise. Likewise, the bottom right panel shows that the TDE is fast-decaying and luminous.

\begin{figure}
    \centering
    \includegraphics[width=0.5\textwidth]{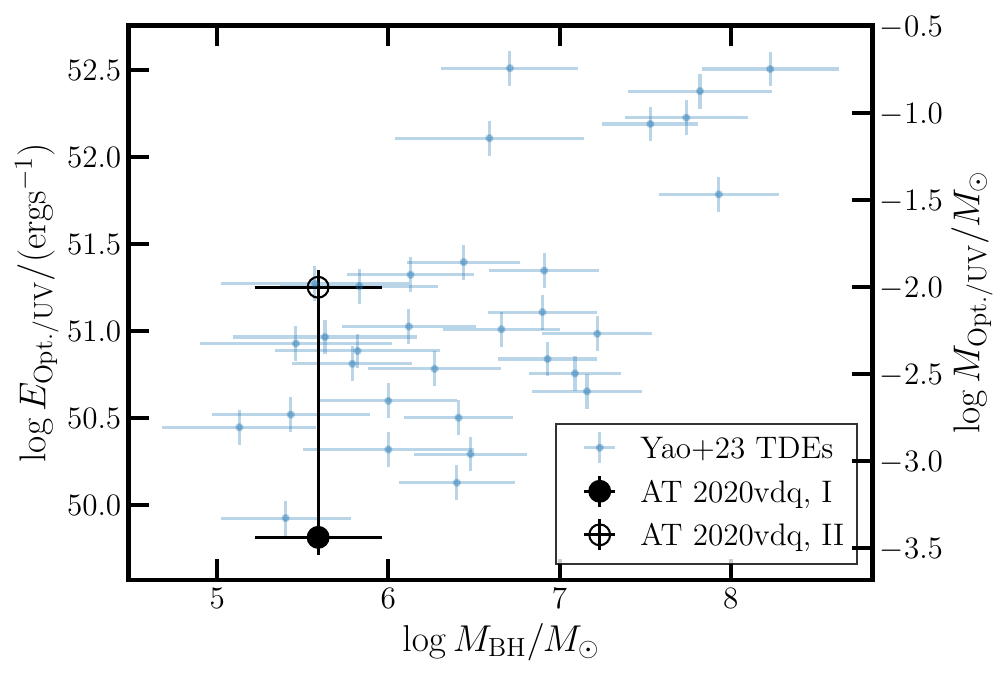}
    \caption{The total emitted optical/UV energy inferred from the pTDE optical/UV flares, in the same format as Figure~\ref{fig:TDE_comp}. The left axis shows the emitted energy while the right axis shows the equivalent stellar mass, assuming an accretion efficiency of $10\%$.}
    \label{fig:Ebb}
\end{figure}
We integrate this bolometric luminosity to $200$ days post-flare to calculate the total energy stored in the outflow. The second flare from AT\,2020vdq was only detected to ${\sim}50$ days, so extrapolating to many hundreds of days would require significant assumptions about the flare shape. It is thus likely that our energy is slightly underestimated. The resulting bolometric energies emitted during the first and second flares are shown in Figure~\ref{fig:Ebb} and tabulated in Table~\ref{tab:optlc}. We also compute the equivalent stellar mass assuming an accretion efficiency of 10\%. We computed that bolometric energies for the \cite{Yao_tdes} following the same procedures. Intriguingly, the energy released during both flares of AT\,2020vdq is comparable to that from the \cite{Yao_tdes}, despite that fact that partial disruptions only involve a fraction of the stellar mass. In fact, the second flare from AT\,2020vdq was among the most energetic TDE optical/UV flares. We will discuss the implications of this result in Section~\ref{sec:discussion}. 

\begin{figure*}
    \centering
    \includegraphics[width=\textwidth]{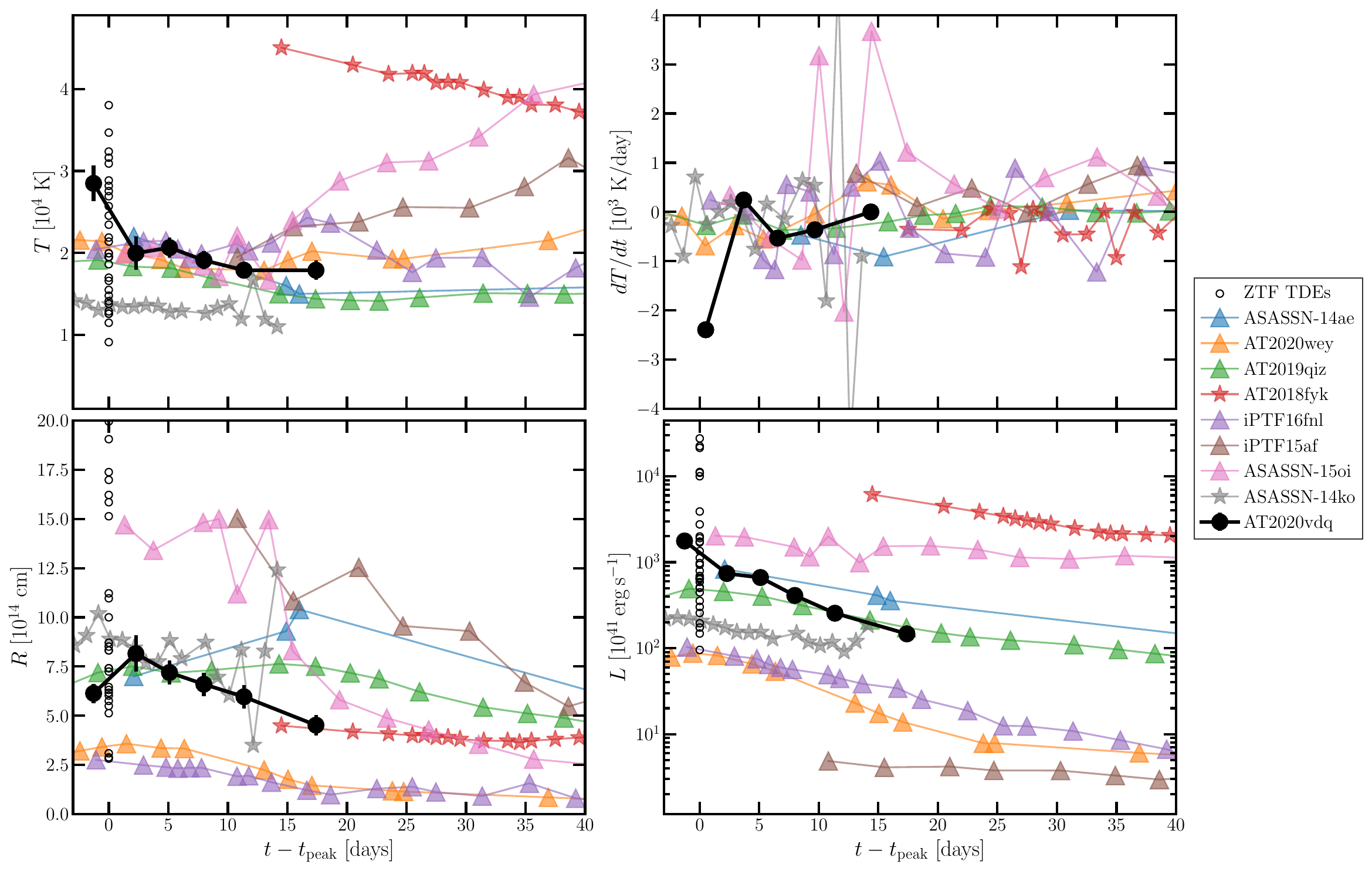}
    \caption{Non-parametric temperature, radius, and luminosity evolution for those TDEs and pTDEs that have available multi-epoch UV and optical data. Data from the rebrightening of AT\,2020vdq is shown as black circles. The required multi-epoch UV data is not available for the initial flare, so only the rebrightening is shown. The candidate pTDEs AT 2018fyk and ASASSN-14ko are shown as colored stars, while normal TDEs are shown as open black circles. The data is retrieved from \cite{14ae, 20wey, qiz,16fnl,iptf15af_stis, 15oi, 14ko_discovery}. AT\,2020vdq shows a significant initial cooling and a rapid luminosity evolution. Otherwise, this pTDE is generally consistent with the normal TDEs.  }
    \label{fig:TRevol}
\end{figure*}

We can improve upon the parametric model fits, which do not provide a good fit near the peak of the rebrightening, by fitting independent blackbodies to the observations at multiple epochs. This is possible because we have multiple epochs of UV observations for the rebrightening, so we can tightly constrain the blackbody parameters for those epochs without relying on a parametric model. We first interpolate the optical lightcurves to the UV observation epochs using a Gaussian process model with a Matern kernel, as implemented in the \texttt{sklearn} package. We then fit the UV+optical for each epoch with a blackbody curve. The resulting temperature, radius, and luminosity evolution is shown in Figure~\ref{fig:TRevol}. As expected given the poor fit near peak from the parametric model, the peak luminosity and temperature from these single-epochs fits are significantly altered. The peak luminosity is $\log L_{\rm bb}/({\rm erg\,s}^{-1}) = 44.2 \pm 0.06$, corresponding to an Eddington ratio $\lambda = 3.5\pm0.5$. The temperature at peak is $\log T_{\rm bb}/{\rm K} = 4.4 \pm 0.03$. The temperature cools dramatically at ${\sim}-2300$ K/day in the first few days of the flare, and then stabilizes to ${\sim}-200$ K/day.

We overlay the results from similar analyses of other fTDEs and pTDEs in Figure~\ref{fig:TRevol}. Candidate pTDEs are distinguished by star markers, while fTDEs are circle markers. Best-fit values at peak are shown for all ZTF TDEs, although we note that these values are measured using parametric fits that do not all include UV observations. The initial temperature evolution of AT\,2020vdq is significantly faster than that of all other TDEs, although this trend is entirely governed by a single data point. Many other TDEs do, however, show some initial cooling within the first ${\sim}20$ days post-peak. The typical cooling rate in these ${\sim}20$ days is comparable to that of AT\,2020vdq if we ignore the first observation. In particular, the temperature evolution from days ${\sim}5-15$ resembles that of AT 2020wey. The temperature is significantly hotter than that observed from the partial TDE ASASSN-14ko. The radius evolution is generally similar to other TDEs. The luminosity evolution, however, is remarkably rapid. This luminosity fades on a timescale comparable to the ``fast and faint'' TDEs AT2020wey and iPTF16fnl despite being an order of magnitude brighter than these events. 

We conclude by briefly considering the flux level between the two flares. We calculated the average flux between ${\sim}300$ days after the first flare to ${\sim}100$ days before the second flare. We found a quiescent $g$-band luminosity $L_g \approx 4 \times 10^{37}$\,erg\,s$^{-1}$ ($14.5\sigma$ significance) and an $r$-band luminosity $L_r \approx 2\times10^{37}$\,erg\,s$^{-1}$ ($7\sigma$ significance). Such emission may be expected if an accretion disk survives between the partial disruptions.

\subsection{Transient spectral optical emission} \label{sec:inter}

\begin{figure*}
    \centering
    \includegraphics[width=\textwidth]{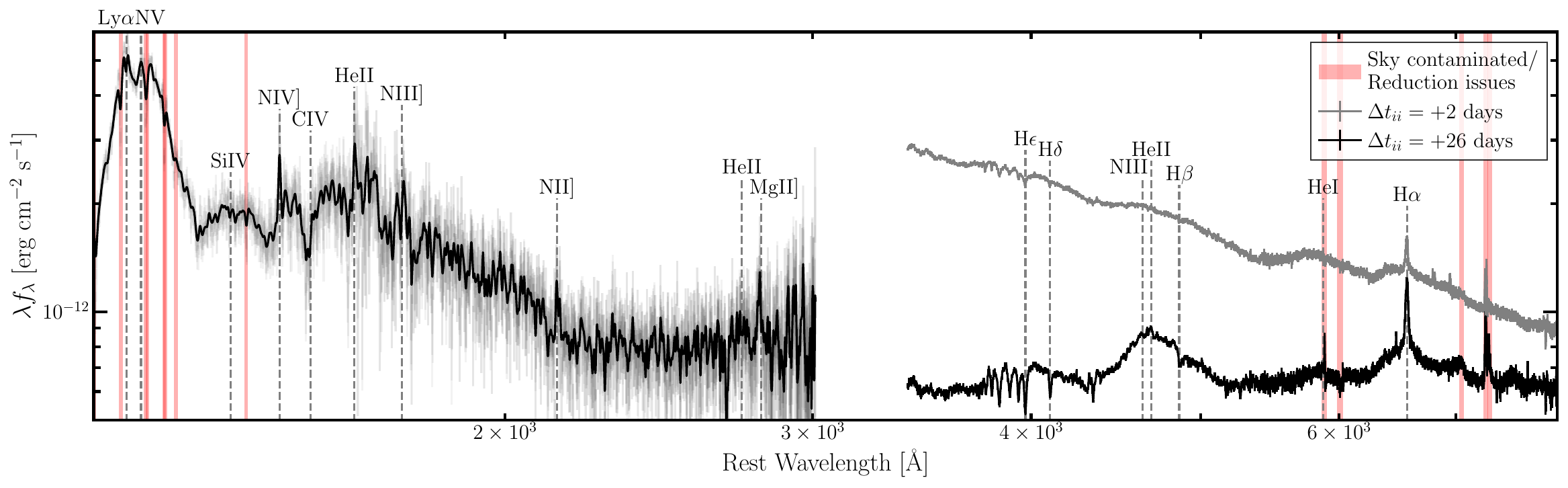}
    \caption{Summary of post-rebrightening UV/optical spectra of AT\,2020vdq. Red regions are contaminated by sky background, telluric lines, or reduction issues. In all spectra, a variety of strong broad, intermediate width, and narrow features are visible on top of a blue continuum. }
    \label{fig:UVopt_spec}
\end{figure*}

We obtained optical spectra of AT\,2020vdq with the LRIS/Keck I telescope using the same settings as \cite{paperII}. These spectra were observed 3 and 27 days post-optical peak (May 21 and Jun. 14 2023; MJDs 60085, 60109). These LRIS spectra are shown in Figure~\ref{fig:UVopt_spec} along with a UV spectrum that will be discussed in the next section. All spectra were reduced following standard procedures.

In both LRIS spectra, the host galaxy features, a blue continuum, and transient spectral features are significantly detected. In particular, we see clear evidence for a broad He\,II$\lambda 4686$/H$\beta$ blend, broad He\,I$\lambda 5876$, and both broad and intermediate-width H$\alpha$. Unfortunately, measuring the broad line fluxes in the spectra is complicated by strong degeneracies between the broad lines and the continuum. We attempt to model the spectra as the sum of broad Gaussian components near the expected wavelengths of He\,II$\lambda 4686$, H$\beta$, He\,I$\lambda 5876$, and H$\alpha$ with a fifth degree Legendre polynomial. We also require an intermediate width H${\rm \alpha}$ component, although we postpone detailed discussion of that line to later in this section, and a Gaussian absorption line near the H$\beta$ wavelength.

\begin{deluxetable}{cc | cc}
\centerwidetable
\tablewidth{10pt}
\tablecaption{ Broad line properties and evolution \label{tab:broad_optlines}}
\tablehead{ \colhead{Param.} & \colhead{} & \colhead{$\Delta t_{\rm ii} = +2$} & \colhead{$\Delta t_{\rm ii} = +26$} \cr 
\colhead{} & \colhead{} & \colhead{(MJD 60085)} & \colhead{(MJD 60109)}}
\startdata
${\rm FWHM}_{\rm b}$ & km\,s$^{-1}$ & $25227\pm112$ & $20038\pm112$\\ 
$\frac{{\rm FWHM}_{\rm b}}{c}$ &  & $0.084\pm0.0$ & $0.067\pm0.0$\\ 
$\Delta v$ & km\,s$^{-1}$ & $-256\pm52$ & $-1225\pm51$\\ 
$L_{\rm H{\rm \alpha}}$ & $10^{40}$\,erg\,s$^{-1}$ & $9.3\pm0.1$ & $5.8\pm0.1$\\ 
$L_{\rm H{\rm \beta}}$ & $10^{40}$\,erg\,s$^{-1}$ & $4.5\pm0.1$ & $2.5\pm0.1$\\ 
$L_{\rm He\,II\lambda 4686}$ & $10^{40}$\,erg\,s$^{-1}$ & $3.1\pm0.1$ & $7.1\pm0.1$\\ 
$L_{\rm He\,I\lambda 5876}$ & $10^{40}$\,erg\,s$^{-1}$ & $7.8\pm0.1$ & $1.4\pm0.1$
\enddata
\tablecomments{Broad line fit parameters. We caution against directly interpreting the reported luminosities as our fit makes many simplified assumptions and does not fully account for the strong degeneracies between the line fluxes and the continuum level/shape.}
\centering
\end{deluxetable}

\begin{figure*}
    \centering
    \includegraphics[width=\textwidth]{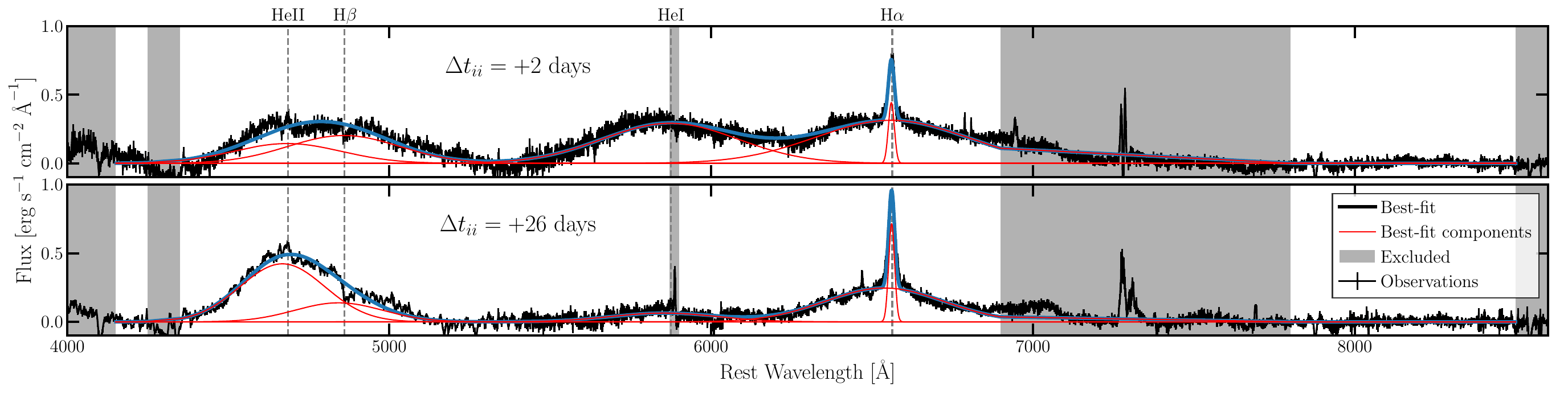}
    \caption{Best-fit models of the broad emission lines from the post-rebrightening, AT\,2020vdq optical spectra. The observed spectra are shown in black. The best-fit models are shown in blue and the best-fit model subcomponents are shown in red. Regions highlighted in gray are excluded from the fit. Broad Balmer, intermediate width Balmer, and Helium lines are detected. The Balmer lines fade slightly between the two spectra while the He\,II line brightens and the He\,I line fades dramatically, suggesting that these lines all originate from different locations in the source. }
    \label{fig:broad_optlines}
\end{figure*}

In Figure~\ref{fig:broad_optlines}, we show the continuum subtracted spectra with the best-fit lines overlaid. The best-fit optical broad line parameters are shown in Table~\ref{tab:broad_optlines}. Unfortunately, the continuum level, which is the sum of galactic emission and the TDE continuum emission, is highly degenerate with the line fluxes (and, to a lesser extent, the line widths) and rendered uncertain by imperfections in the flux calibration.  Moreover, we have fixed the FWHM and velocity offsets of all the broad lines to a single value to remove degeneracies between the broad He\,II and H$\beta$ emission; however, these assumptions are not physically motivated and likely cause erroneous flux measurements. Hence, we caution against detailed interpretation of the line luminosities. The FWHM measurement is likely more reliable, although it is still subject to significant degeneracies. Given the aforementioned uncertainties, we will not discuss and interpret the line fluxes in detail, other than to note that the Balmer decrement may be tantalizingly flat (H${\rm \alpha}/$H${\rm \beta} \approx 2$), as has been observed for previous TDEs \citep[][]{hyz_spectral}. 

The line FWHM are nearly ${\sim}0.1c$ with small blueshifts. The line width narrows slightly with time, while the blueshift increases. The H${\rm \alpha}$, H${\rm \beta}$ and He\,I lines evolve similarly: they begin bright and broad, but fade by the second observation. The He\,II line, in contrast, appears to brighten between the two observations, suggesting an alternate origin of the line. Similar behavior has been reported from the TDE candidate AT 2018hyz \citep[][]{hyz_spectral}.

If we assume that the line widths correspond to the velocity of outflows launched during the pTDE, that the optical broadband emission is produced by the same outflowing gas, and that the electron scattering optical depth is $\tau_{\rm es} \approx 1$, we can calculate the amount of material launched in outflows. The mass in the outflow $M_{\rm outflow}$ is given by 
\begin{equation}
    M_{\rm outflow} = \dot{M} \Delta t = 4\pi \rho v R^2 \Delta t.
\end{equation}
Here, $\rho$ is the density of the outflow, $v$ is the velocity of the outflow, $R$ is the radius at which the optical emission is produced, and $\Delta t$ is the duration of the outflow. Given the rapid drop in the optical emission after ${\sim}15$ days, we expect that most of the outflows were launched early in the flare and we adopt $\Delta t \approx 15$ days. The average radius in this period is $R \sim 10^{14.8}$ cm. Assuming a fully ionized Hydrogen and Helium plasma, for $\tau_{\rm es} = 1$ we have $\rho R = \kappa_{\rm es} = 0.34$ g cm$^{-2}$. Then,
we find $M_{\rm outflow} \approx 0.05\,M_\odot$.

If the electron scattering optical depth is large, line broadening in TDEs may instead be dominated by electron scattering \citep[][]{es_lines}. For a moderate optical depth, a narrow line is produced by the unscattered photons and a broad base is visible due to the scatter photons. While we do observe both broad and narrow line emission in our spectra, it seems unlikely that the narrow line emission is the unscattered component of the emission: He\,I and H$\alpha$ are at similar wavelength and have similar broad line fluxes in the early spectrum, yet H$\alpha$ shows a strong narrow line and no detectable narrow He\,I line is present. Unless the He\,I emission region has a substantially higher optical depth than the H$\alpha$ emitting region, the narrow lines cannot be produced by the same mechanism as the broad lines. Hence, we assume that no significant unscattered component is detectable, so the emitting regions all have large electron scattering optical depths. It is beyond the scope of this work to calculate the exact optical depth required to produce the line widths observed, although we note that none of the line profiles presented by \cite{Roth2018} are as broad as the lines observed here. We urge attempts to reproduce the spectrum observed with this source with electron scattering models to determine if it is possible to match our observations. Based on Figure 3 of \cite{Roth2018}, symmetric broad lines similar if not as broad as those produced by AT\,2020vdq are emitted by static or near-static gas with $\tau_{\rm es}\gtrsim 8$. Adopting $R\sim 10^{14.8}$ cm as before, this corresponds to a gas density $\rho \gtrsim 4\times 10^{-15}$ g cm$^{-3}$ or a number density $n \gtrsim 10^{9}$ cm$^{-3}$. The total mass in this outflowing gas is thus $0.002\,M_\odot$.

\begin{figure*}
    \centering
    \includegraphics[width=\textwidth]{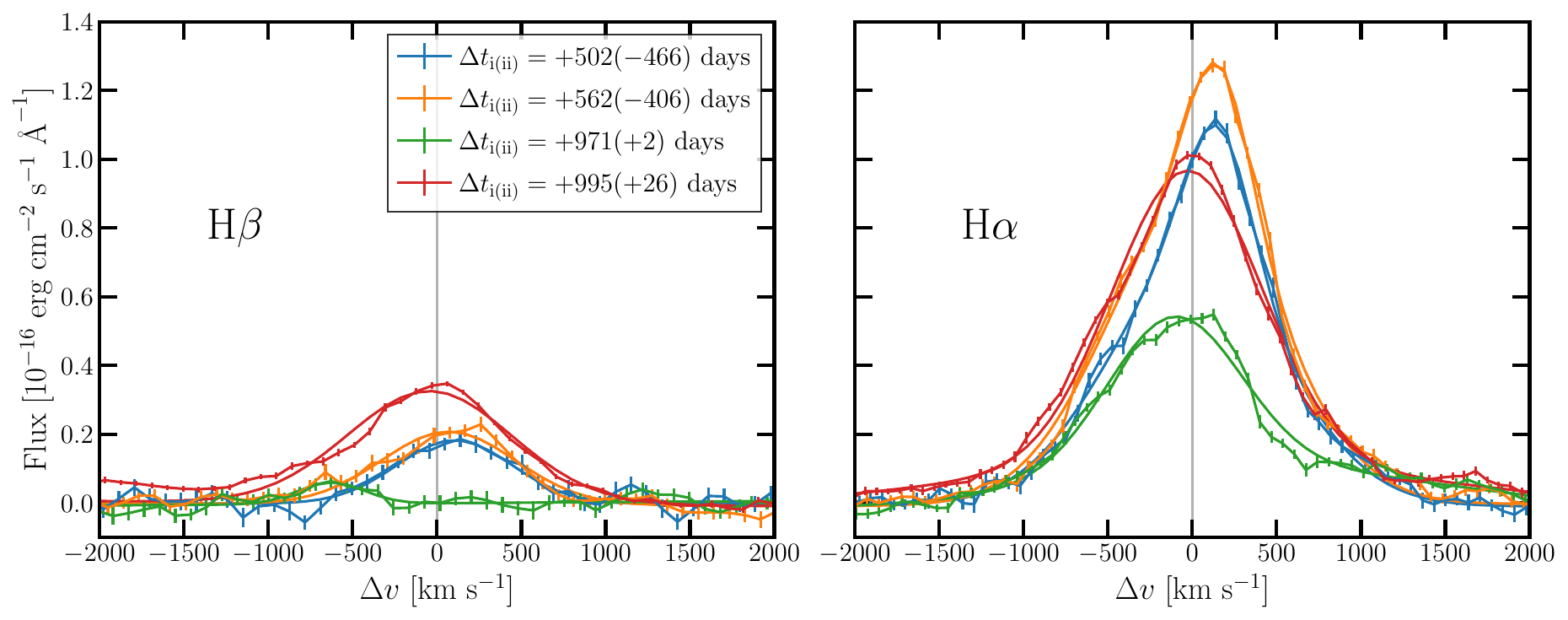}
    \caption{The evolution of the intermediate Balmer lines observed from AT\,2020vdq. H$\beta$ is shown in the left panel while H$\alpha$ is shown in the right. Data is shown as lines with errorbars while Gaussian model fits are shown as lines. The Balmer lines were brightening ${\sim}2$ years after the first flare, but faded and became slightly blueshiftedby the rebrightening. A few weeks post-rebrightening, the Balmer lines began brightening while remaining blueshifted.}
    \label{fig:balmer}
\end{figure*}

\begin{deluxetable*}{cc | cccc}
\centerwidetable
\tablewidth{10pt}
\tablecaption{ Intermediate Balmer line properties and evolution \label{tab:balmer}}
\tablehead{ \colhead{Parameter} & \colhead{Unit} & \colhead{$\Delta t_{\rm ii} = -466$ days} & \colhead{$\Delta t_{\rm ii} = -406$ days} & \colhead{$\Delta t_{\rm ii} = +2$ days} & \colhead{$\Delta t_{\rm ii} = +26$ days} \cr
\colhead{} & \colhead{} & \colhead{(MJD 59616)} & \colhead{(MJD 59676)} & \colhead{(MJD 60085)} & \colhead{(MJD 60109)}}
\startdata
$L_\alpha$ & $10^{40}\,$erg$\,$s$^{-1}$ & $1.231 \pm 0.013$ & $1.452 \pm 0.011$ & $0.874 \pm 0.022$ & $1.69 \pm 0.05$\\ 
${\rm FWHM}_\alpha$ & km\,s$^{-1}$ & $855 \pm 18$ & $892 \pm 12$ & $1077 \pm 18$ & $1152 \pm 10$\\ 
$\Delta v_\alpha$ & km\,s$^{-1}$ & $128 \pm 6$ & $128 \pm 5$ & $-94 \pm 6$ & $-26 \pm 2$\\ 
$L_\beta$ & $10^{40}\,$erg$\,$s$^{-1}$ & $0.128 \pm 0.009$ & $0.171 \pm 0.008$ & $0.025 \pm 0.005$ & $0.323 \pm 0.004$\\ 
${\rm FWHM}_\beta$ & km\,s$^{-1}$ & $807 \pm 57$ & $932 \pm 47$ & $464 \pm 99$ & $1145 \pm 14$\\ 
$\Delta v_\beta$ & km\,s$^{-1}$ & $-104 \pm 22$ & $-70 \pm 19$ & $630 \pm 42$ & $42 \pm 6$\\ 
${\rm L_\alpha}/{\rm L_\beta}$ &  & $9.6 \pm 0.7$ & $8.5 \pm 0.4$ & $35 \pm 8$ & $5.22 \pm 0.17$\\ 
\enddata
\tablecomments{Best-fit Gaussian model parameters to the intermediate width Balmer lines, as described in Section~\ref{sec:inter}.}
\centering
\end{deluxetable*}

In addition to these broad lines, we observe strong intermediate-width H$\alpha$ emission, similar to that observed following the first flare. These lines are treated in detail in \cite{paperII}, so we only briefly discuss the properties and interpretation here. The evolution of these intermediate width lines from before to after the rebrightening is shown in Figure~\ref{fig:balmer}. We have removed the stellar continuum using a fit with the \texttt{ppxf} software. Following \cite{paperII}, we found that the H$\alpha$ lines could be well modeled as the sum of two Gaussians. The H$\beta$ lines are well-modeled as single Gaussians. Those fits are overlaid in Figure~\ref{fig:balmer} and the best-fit parameters are summarized in Table~\ref{tab:balmer}. 

Before the rebrightening, the emission lines were peaked near the host redshift and were brightening. They began fading at some point in the ${\sim}400$ days before the source rebrightened. In the earliest spectrum post-rebrightening, the lines were slightly blueshifted relative to the host. They had brightened by the next spectrum, ${\sim}24$ later or $27$ days post-peak, suggesting that the emitting material is located ${\lesssim}7\times10^{16}$ cm from the SMBH. Assuming the emission is produced via recombination and using this distance upper limit as an upper limit on the size of the emitting region, we find that the gas density is $\gtrsim 10^5$ cm$^{-3}$. Such a high density suggests that we are seeing emission from very compact clumps of outflowing gas.

\subsection{UV spectrum}

\begin{figure*}
    \centering
    \includegraphics[width=\textwidth]{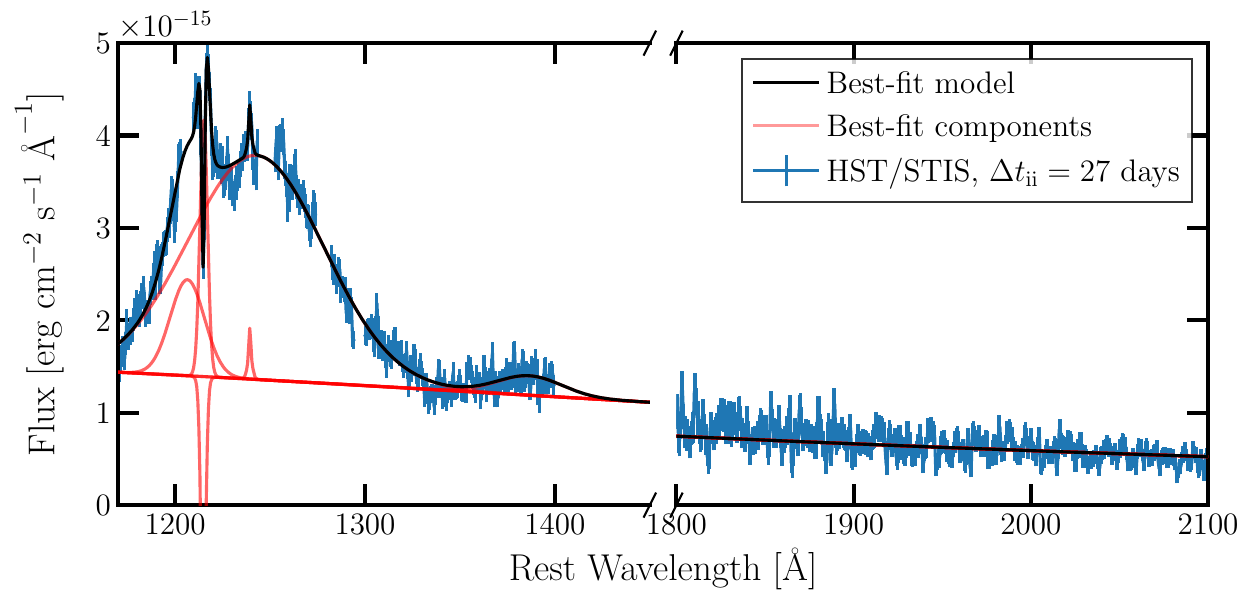}
    \caption{Fit to the Ly$\alpha$ region and a continuum region of the {\it HST} spectrum, in the same format as Figure~\ref{fig:broad_optlines}. The spectrum is well-modelled as two broad lines (Ly$\alpha$ and an ambiguous line near Si\,IV$\lambda 1394$), one intermediate line (Ly$\alpha$), two narrow emission lines (Ly$\alpha$ and N\,V), and one narrow absorption line (Ly$\alpha$). The broad Ly$\alpha$ line is redshifted by ${\sim}2\times 10^4$\,km\,s$^{-1}$ relative to the host rest-frame, which is in strong contrast to the optical broad spectral features.}
    \label{fig:HST_fit}
\end{figure*}

We obtained a UV spectrum on June 14, 2023 (MJD 60109) using the Space Telescope Imaging Spectrograph (STIS) on the {\it Hubble Space Telescope} ({\it HST}). We observed AT\,2020vdq with the G140L grating and the FUV-MAMA detector for 3560 seconds and with the G230L grating and the NUV-MAMA detector for 1082 seconds, for a total of two orbits. All observations used the $52\times0.2\arcsec$ aperture. The resulting UV spectrum is shown in Figure~\ref{fig:UVopt_spec}.

We begin with a qualitative description of the spectrum. Strong, hot blackbody emission is detected in the continuum of the UV spectrum. On top of this continuum, a number of emission lines are detected. First, an extremely broad Ly${\rm \alpha}$ emission line is present, which is slightly redshifted and may have a flat top. Superposed on this broad line is an intermediate width Ly$\alpha$ emission line with a narrow absorption component. There are other broad features near the wavelengths of Si\,IV$\lambda 1394$, N\,IV$]\lambda 1486$, and C\,IV$\lambda \lambda 1548,1551$. Near but blueshifted relative to the wavelength of C\,IV, two narrow absorption lines are present. A number of narrow emission lines are also detection: N\,IV$]\lambda 1486$ and Mg\,II]$\lambda\lambda 2796,2803$ are the most significant and He\,II$\lambda 1640$, N\,III]$\lambda\lambda 1747,1749$, and N\,II]$\lambda 2143$ are likely detected. These lines are often detected in galaxies, so we assume they are associated with the host. We do not have sufficient signal-to-noise to perform a detailed analysis of the line emitting regions, so we do not discuss these lines further.

In the rest of this section, we attempt to quantitatively analyze and interpret these features. We focus on the broad Ly$\alpha$. While the other broad emission lines may be present, our spectrum is not of sufficiently high signal-to-noise to characterize them.

\begin{deluxetable}{c | ccc}
\centerwidetable
\tablewidth{10pt}
\tablecaption{ UV spectral features \label{tab:UV}}
\tablehead{ \colhead{Name} & \colhead{Lum.} & \colhead{FWHM} & \colhead{$\Delta v$} \cr
\colhead{} & \colhead{$10^{40}$ erg s$^{-1}$} & \colhead{km s$^{-1}$} & \colhead{km s$^{-1}$}}
\startdata
Ly$\alpha_1$ & $95 \pm 5$ & $18747 \pm 646$ & $7556 \pm 344$ \cr 
Ly$\alpha_2$ & $16 \pm 3$ & $7290 \pm 650$ & $-2883 \pm 252$ \cr 
SiIV & $3 \pm 1$ & $7830 \pm 2086$ & $-1401 \pm 758$ \cr
\enddata
\centering
\end{deluxetable}

We begin with the broad Ly$\alpha$ line. We fit this line by modelling the region from $1200{-}1400\,{\rm \AA}$ as the sum of Gaussians and a blackbody. We also include the continuum-dominated region from $1800{-}2100\,{\rm \AA}$ in the fit to help constrain the blackbody parameters. While redder regions of the spectra also are continuum-dominated, we exclude them because the galaxy contribution is expected to be larger and potentially dominant. Seven Gaussians are required for a reasonable fit: (1) a Ly$\alpha$ broad component, (2) an intermediate-width Ly$\alpha$ component, (3) a broad component near $1400\,{\rm \AA}$ (possible Si\,IV$\lambda 1394$) (4) a narrow Ly$\alpha$ absorption line, and (5-6) two narrow Ly$\alpha$ and N\,V$\lambda 1239$ emission components. We generally adopt broad bounds for the parameters in this fit, including for the blackbody temperature and radius. The best fit with subcomponents is shown in Figure~\ref{fig:HST_fit} and the best-fit broad/intermediate line parameters are tabulated in Table~\ref{tab:UV}.

Like the fit to the broad optical lines, this fit is subject to strong degeneracies between the continuum and line parameters, so we do not attempt detailed interpretation. We focus on the intermediate-broad Ly$\alpha$ parameters. The broad L$\alpha$ is well-modelled as a Gaussian redshifted by ${\sim}7600$ km s$^{-1}$ and with a FWHM ${\sim}19000$\,km\,s$^{-1} = 0.06c$. This FWHM is consistent with those of the optical broad lines; however, the optical lines are slightly blueshifted whereas the Ly$\alpha$ line is redshifted. It is possible that a Gaussian is not the correct parameterization for this line, which could affect the velocity offset. If this line is, e.g., flat-topped or has multiple components (both of which can be produced by accretion disks and have been observed from past TDEs; \citealp{hyz_spectral}), the redshift may be lessened. However, regardless of the model, it is clear from inspection that this broad line will always be at least somewhat redshifted, so we expect that it is coming from a different outflow component from the optical broad line emission. Based on the line width, all the broad lines may be produced at a similar distances from the black hole.

We require a second, intermediate component in our Ly$\alpha$ model that has a FWHM${\sim}7000$\,km\,s$^{-1}$ and a blueshift of $\Delta v \sim 3000$ km s$^{-1}$. This line is broader than the intermediate width components observed in the optical. However, the measured width should be interpreted with caution because there is a strong sky line near the rest-frame Ly$\alpha$ and changing the parameterization of the broad Ly$\alpha$ line would likely significantly affect the parameters of this line. Hence, we cautiously regard this as a analogous to the intermediate width lines in the optical, despite the FWHM differences.

A few past TDEs (both fTDEs and pTDEs) have UV spectroscopy; a good summary of observations is provided by Figure 11 of \cite{14ko_stis}. Broad Ly$\alpha$ lines, like that observed here, are detected in most near-peak TDE UV spectra \citep[][]{14ko_stis,16fnl_stis,14li_stis,2018zr_stis,qiz_stis}, although the emission from AT\,2020vdq appears significantly broader than typically observed (note that few authors provide fits to the broad emission, so a quantitative comparison is not possible). iPTF 15af shows a similarly broad line, although with strong absorption features superimposed that we do not see here \citep[][]{iptf15af_stis}. The broad Si\,IV and C\,IV are also frequently observed \citep[e.g.][]{14ko_stis}. Intermediate width emission from Ly$\alpha$ has not been reported before, to our knowledge. Thus, other than the intermediate width components and the extreme width of the broad emission, the UV spectrum of AT\,2020vdq resembles those of past TDEs.

\subsection{Radio emission}

We obtained radio follow-up shortly after the rebrightening of AT\,2020vdq using the VLA telescope (${\sim}1-20$ GHz; Proposal AS1800, PI Somalwar) and the NOEMA telescope (mm; proposal E22AH, PI Somalwar). All observations used standard configurations. We observed AT\,2020vdq with the VLA in the S, C, X, K, and Ku bands on May 26, 2023 (MJD 60090), or $7$ days after the peak of the optical/UV rebrightening. The VLA observations were reduced using the most recent CASA VLA pipeline (2022.2.0.64). The source was detected in all bands and was sufficiently bright in S, C, and X to image individual spectral windows.  The signal-to-noise was lower in the K and Ku bands, so we imaged all spectral windows together.

We observed AT\,2020vdq at 95.7 GHz with the NOEMA telescope on June 1, 2023 (MJD 60096), or $14$ days after the rebrightening peak. The observations were reduced using recommended pipeline procedures. The source was not detected.

The resulting fluxes and upper limits are shown in Figure~\ref{fig:radio_sed}. The fluxes can be well-modeled as a single power-law: there is no new outflow component detected. 

\begin{deluxetable}{cc}
\centerwidetable
\tablewidth{10pt}
\tablecaption{ Swift/XRT $5\sigma$ upper limits \label{tab:swift}}
\tablehead{ \colhead{MJD} & \colhead{AT\,2020vdq} \cr
\colhead{} & \colhead{[$10^{40}$ erg s$^{-1}$]}}
\startdata
59638.2 & ${<}34.6$ \cr
60081.2 & ${<}6.2$ \cr
60084.8 & ${<}11.4$ \cr
60087.3 & ${<}96.5$ \cr
60090.5 & ${<}90.8$ \cr
60093.8 & ${<}90.8$ \cr
60099.8 & ${<}4.4$ \cr
60102.2 & ${<}147.6$ \cr
60105.5 & ${<}170.3$ \cr
60108.1 & ${<}4.0$ \cr
\enddata
\centering
\end{deluxetable}

\begin{figure}
    \centering
    \includegraphics[width=\columnwidth]{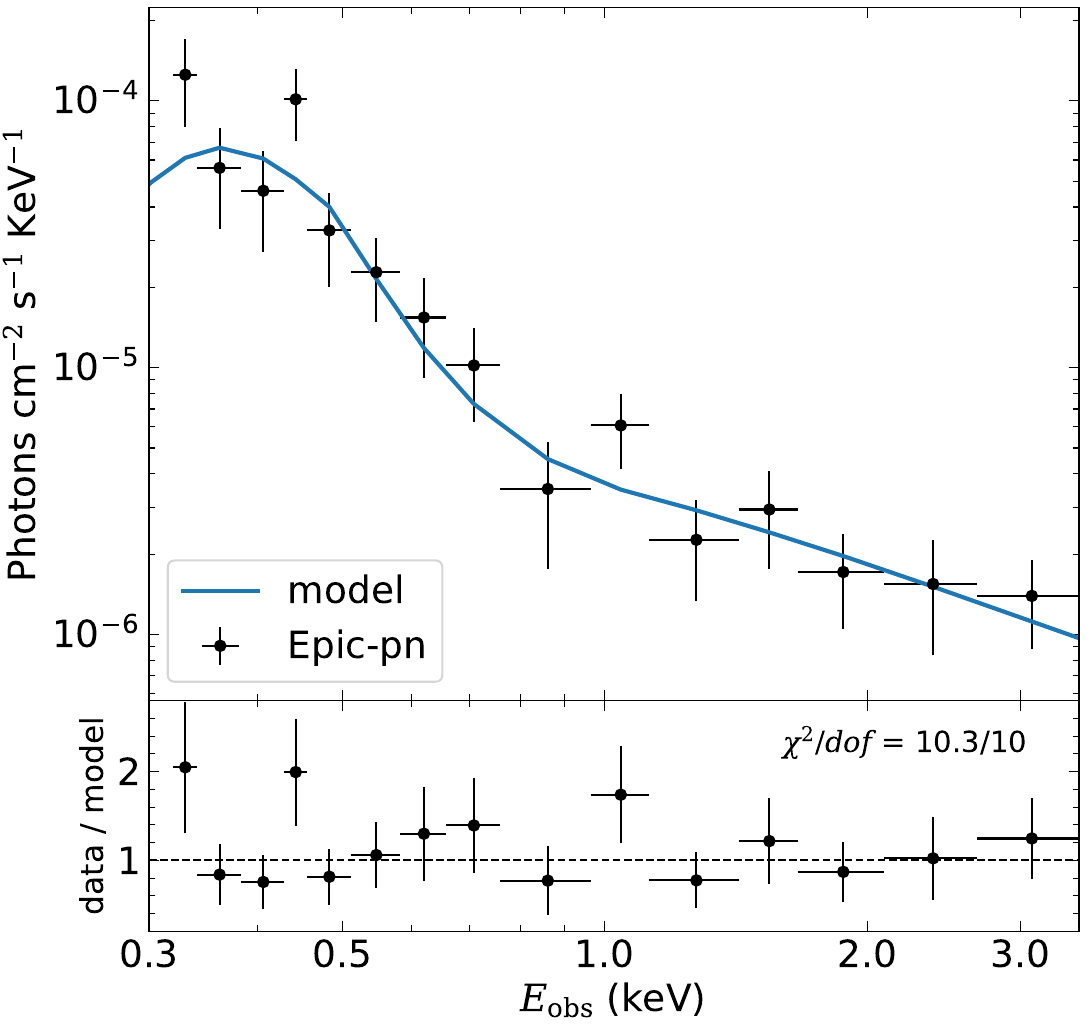}
    \caption{{\it XMM-Newton} spectrum of AT\,2020vdq, observed $+2$ days post the peak of the rebrightening. The data is shown as black points while the best-fit \texttt{TBabs$\times$zashift$\times$(simpl$\otimes$diskbb)} model is shown as as a blue line.}
    \label{fig:xmm}
\end{figure}

\subsection{X-ray emission}
\newcommand\xmm{\textit{XMM-Newton}\xspace}
We (among other groups) obtained X-ray follow-up of AT\,2020vdq during the rebrightening using Swift/XRT (PIs: Lin, Leloudas, Guolo) and XMM-Newton (PI Somalwar). The Swift/XRT observations were performed in photon counting mode and reduced using standard \texttt{heasoft} procedures. We used the \texttt{sosta} function to perform forced photometry at the location of the source using default parameters. The source was not detected in any single Swift/XRT epoch; upper limits on shown in Table~\ref{tab:swift}. 

AT\,2020vdq was observed and detected by XMM-Newton in ${\sim}10$ ks on MJD 60085 ($+2$ days post peak; May 21, 2023). The observation data files (ODFs) were reduced using the \xmm Standard Analysis Software \citep[SAS;][]{Gabriel_04}, and the detailed reduction processes as described in \citet{Guolo_xray} were followed.
The source spectrum, shown in Figure~\ref{fig:xmm}, is detected above the background up to $\sim$ 3.0 keV. The following spectral fitting procedures were done using the python version of \texttt{xspec} \citep{Arnaud1996}, \texttt{pyxspec}\footnote{https://heasarc.gsfc.nasa.gov/xanadu/xspec/python}.
For all spectral models described below, we included 
the Galactic absorption using the \texttt{TBabs} model \citep{Wilms2000}, with the hydrogen-equivalent neutral column density $N_{H}$ fixed at the galactic value $N_{\rm H} = 1.38 \times 10^{20}$ cm$^{-2}$. We shifted the TDE emission using the convolution model \texttt{zashift}. We start by trying to model the spectrum with a phenomenological powerlaw (\texttt{TBabs$\times$zashift$\times$powerflaw}), as expected \cite{Guolo_xray}, the powerlaw model is an inadequate model for TDE X-ray emission, the resulting $\chi^2$/degrees of freedom ($dof$) $= 20/12$. Alternatively, we also employ a purely thermal model (\texttt{TBabs$\times$zashift$\times$diskbb)}), which also results in a poor fit  $\chi^2/dof = 30/12$, from a strong residual at energies higher than 1.0 keV. 

Finally, based on \citet{Guolo_xray}, we combined the thermal model with a convolution model (\texttt{simpl}, \citealp{Steiner_09}) that 
emulates the comptonization process to create a powerlaw. With two free parameters: $f_{\rm sc}$, the fraction of comptonized photons, 
and $\Gamma$, the resulting powerlaw photon index. Our final model \texttt{TBabs$\times$zashift$\times$(simpl$\otimes$diskbb)}, 
results in a great fit $\chi^2/dof = 10.3/10$, the best-fitted parameters are shown Table~\ref{tab:xmm}. The unabsorbed $0.3{-}10$\,keV 
flux was $\log f_X/({\rm erg\,cm}^{-2}\,{\rm s}^{-1}) = -12.95 \pm 0.05$, or $\log L_X/({\rm erg\,s}^{-1}) = 41.4 \pm 0.05$ ($L_X = 
0.005 L_{\rm edd.}$)

This X-ray emission is consistent with both observations of X-ray faint TDEs shortly post-peak \citep[][]{xraytde_review,Guolo_xray} 
and late-time X-ray observations of TDEs \citep[][]{late_xray_tde,Guolo_xray}, so it could be associated with the new flare or it 
could be relic emission from the accretion disk created during the first flare. The temperature of the thermal component is consistent 
with other optically selected TDEs \cite{Guolo_xray}, and the faint powerlaw emission ($f_{\rm sc} = 0.03^{+0.01}_{-0.01}$), could 
originate both from the reprocessing layer that produces the bright UV/optical, or a nascent corona \citep[][]{Thomsen2022,Guolo_xray}, late 
time observations may be able to distinguish them.

\begin{deluxetable}{ccc}
\centerwidetable
\tablewidth{10pt}
\tablecaption{Best-fitted  parameters \xmm spectrum. \label{tab:xmm}}
\tablehead{ \colhead{Model} & \colhead{Component }  & \colhead{Value}}
\startdata
\texttt{TBabs}                   & $N_{\rm H}$  &     $1.38\times10^{20}$ cm$^{-2}$ (fixed)  \\
\texttt{zashift}                 & $z$          &    0.044 (fixed)   \\
\multirow{2}{*}{\texttt{simpl}}  & $f_{\rm sc}$ &     $0.03^{+0.01}_{-0.01}$  \\
                                 & $\Gamma$     &      $1.35^{+0.31}_{-0.25}$ \\
\multirow{2}{*}{\texttt{diskbb}} & $T_{in}$     &      $60^{+6}_{-4}$ eV \\
                                 & norm         &  $820^{+770}_{-250}$   
\enddata
\centering
\end{deluxetable}

\section{Constraints on previous flares from AT\,2020vdq} \label{sec:prev_flares}

\begin{figure*}
    \centering
    \includegraphics[width=\textwidth]{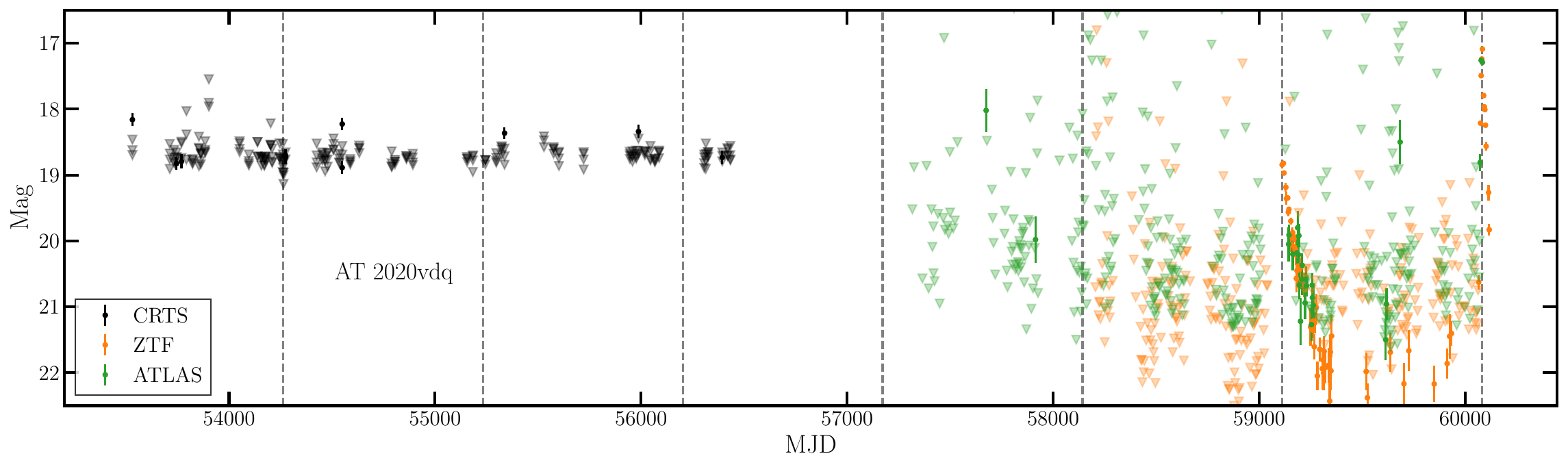}
    \caption{Historical lightcurves for AT\,2020vdq. The lightcurves for each survey/band have been binned in $3$ days bins. For clarity, we distinguish data from different surveys with different colors, but do not show different colors for each band (e.g., the ZTF lightcurve includes $gri$ observations). We find no evidence for earlier flares in any of the data. ATLAS data rule out similar-brightness flares in the last ${\sim}10$ years.}
    \label{fig:hist_lc}
\end{figure*}

In this section, we search for past flares from AT\,2020vdq. We retrieve photometry of this source from the ZTF, the Catalina Real Time Transient Survey (CRTS; \citealp{crts}), and Asteroid Terrestrial-impact Last Alert System  (ATLAS; \citealp{atlas}). As shown in Figure~\ref{fig:hist_lc}, no significant flare is detected by any survey. In both cases, ATLAS coverage is sufficient to rule out previous flares as bright or brighter than the detected flares. We cannot, however, rule out previous flares that had peak magnitudes that were fainter by ${\sim}1$ mag.

\section{Discussion} \label{sec:discussion}

We have presented the ZTF-discovered partial TDE candidate AT\,2020vdq. Key properties of this event is summarized in Table~\ref{tab:summary}.

\begin{deluxetable*}{p{0.2\linewidth} p{0.4\linewidth} p{0.4\linewidth}}
\centerwidetable
\tablewidth{10pt}
\tablecaption{ Summary of observations of AT\,2020vdq \label{tab:summary}}
\tablehead{\colhead{Parameter} & \colhead{Flare I} & \colhead{Flare II}}
\startdata
Host galaxy & \multicolumn{2}{c}{post-starburst, green-valley; $M_{\rm BH}=10^{5.6}\,M_\odot$} \cr
Optical/UV flare & $L_{\rm bb}{=}10^{42.8}\,{\rm erg\,s}^{-1} {=} 0.1L_{\rm edd}$\newline $T_{\rm bb} = 10^4$\,K\newline rose over ${\sim}8$ days\newline decayed over ${\sim}60$ days & $L_{\rm bb} = 10^{44}\,{\rm erg\,s}^{-1} {=} 2L_{\rm edd}$\newline
$T_{\rm bb} = 10^{4.3}$\,K\newline
rose over ${\sim}5$ days\newline
decayed over ${\sim}6$ days\newline
(one of the fastest evolving TDE flares)\cr
Broad lines & no early time spectra & extremely broad (${\sim}0.1c$), mildly blueshifted Balmer, He I, He II, and Ly$\alpha$ lines (H+He TDE) \cr
Narrow lines & ${\sim}1000$\,km\,s$^{-1}$, slightly redshifted (${\sim}100$\,km\,s$^{-1}$) Balmer (with large Balmer decrement), He\,I, He\,II, and Fe\,X detected & ${\sim} 1000$\,km\,s$^{-1}$ lines detected that brightened over the first few weeks, high but variable Balmer decrement \cr
X-ray & No early-time X-ray observations, $L_X \lesssim 3\times 10^{41}$ erg\,s$^{-1}$ ${\sim}1.4$ years post-flare & $L_X {\sim} $ within $2$ weeks post-peak \cr
Radio & Detected as a $L_R \approx 10^{38}$\,erg\,s$^{-1}$ radio transient ${\sim}1.4$ years after the initial flare. The radio-emitting region was consistent with a non-relativistic ($\beta \sim 0.1$), wide-angle outflow. & No new radio-emitting region was detected in the ${\sim}$month post-rebrightening 
\enddata
\centering
\end{deluxetable*}

In this section, we will first argue that AT\,2020vdq is caused by a pTDE rather than a different type of activity. Then, we will discuss AT\,2020vdq in the context of other pTDEs and we will use it to constrain both TDE and pTDE models. 

\subsection{The cause of AT\,2020vdq} \label{sec:comp}

We first argue that AT\,2020vdq is a pTDE. We consider four possible origins of this event (1) a TDE followed by a supernova,  (2) AGN variability, (3) a TDE by a SMBH binary, (4) two independent TDEs, or (5) a repeating partial TDE. 

The first scenario is disfavored by many pieces of evidence. First, the intermediate width Balmer lines were initially caused by a TDE. They began evolving after the second flare, which would not be expected if the second flare is caused by a SN rather than an event directly in the vicinity of the SMBH. In addition, the broad optical spectral features are well-modelled as a TDE, but are less typical for SN. In the case of a Type II SN, the lack of star-formation provides additional evidence against a SN flare origin. Hence, we do not believe that the second flare from AT\,2020vdq could be caused by a SN.

Instead, it is possible that both flares from AT\,2020vdq are related to a flaring AGN. As we argue in \cite{paperII}, the lack of any signatures of AGN activity  from this source render this hypothesis unlikely: the WISE W1-W2 color is $0.088$ mag, which is outside of the AGN regime, and we do not see significant infrared variability pre-flaring. We do not see evidence for strong narrow Balmer on top of the intermediate width transient lines and only weak, narrow [O\,III] is detected. There are also no archival radio nor X-ray detections. AT\,2020vdq could, at most, be an extremely low luminosity AGN. A single optical flare like those from AT\,2020vdq from such an AGN, let alone two, is unprecedented. 

A TDE by a SMBH in a binary is expected to have an accretion rate that initially follows standard predictions for TDE evolution. The accretion rate will then cut off after a fraction of the period of the SMBH binary, but it eventually may increase again \citep[][]{2009ApJ...706L.133L, 2016MNRAS.458.1712R}. In no model of such events, however, is the accretion rate expected to be increase after the first flare. Hence, we would not expect a flare such as that observed from AT\,2020vdq, which is more energetic during the rebrightening.

We next consider the possibility that AT\,2020vdq is two independent TDEs in a single galaxy. The TDE rate has been measured to be ${\sim}3.2 \times 10^{-5}$ yr$^{-1}$ galaxy$^{-1}$ using $33$ ZTF TDEs. After monitoring these TDEs for ${\sim}3$ years post-discovery, we would expect to detect a mean of ${\sim}3.2 \times 10^{-5}$ yr$^{-1}$ Galaxy$^{-1}\times 33$ Galaxies$\times 3$ years$=0.003$ additional TDEs. Thus, the probability of observing one or more additional, independent TDE from these $33$ TDE hosts is $0.3\%$. While this probability is not negligible, it suggests that it is unlikely that we have observed two independent TDEs. However, we urge careful consideration of this possibility in future repeating pTDE analyses, as the probability of observing two independent events in a single galaxy will only increase with time and as TDE samples grow.

Considering that post-starburst galaxies have elevated TDE rates (by a factor of 10 to 100) as compared to other types of galaxies, the time interval between two independent TDEs may be as short as a few hundred years \citep{Arcavi14_TDE_host, french16_host_galaxies, law-smith17_TDE_host, hammerstein21_TDE_host}. For an extreme case of an average rate of $3\times 10^{-3}\rm\, yr^{-1}\, galaxy^{-1}$, the probability of detecting two independent TDEs from $33$ TDE hosts in 3 years is as large as $30\%$. This possibility can be tested by future monitoring of AT\,2020vdq; (not) detecting another flare from AT\,2020vdq in the near future will rule out (or support) this possibility.

We conclude that AT\,2020vdq is most likely a pTDE.

\subsection{AT\,2020vdq in the context of published repeating pTDE candidates} \label{sec:comp}

There are a handful of published partial TDE candidates. We only consider those that flared at least twice. There have been optical/UV pTDE candidates identified based on their low-luminosity and/or fast-evolving flares \citep[e.g.][]{hyz_ptde}; however, non-repeating pTDEs are very difficult to confirm as it may be possible to generate low-luminosity/fast flares through additional mechanisms. The repeating pTDEs that we consider to be reliable candidates are summarized as follows.\\

\noindent {\it ASASSN-14ko \citep[][]{14ko_discovery,14ko_xrayUV,14ko_stis,14ko_followup}: } ASASSN-14ko is one of the most well-studied pTDE candidates discovered to date. It is a nuclear transient in an AGN that was discovered by the ASASSN survey. It flared seventeen times between 2014 and 2020 in the optical, with a period of $114.2$ days. The optical flaring was accompanied by UV flaring and the best-fit flare blackbody parameters and evolution are consistent with typical TDEs (grey stars in Figure~\ref{fig:TRevol}). At least a subset of the optical flares are accompanied by X-ray variability. The optical spectra for this source show possible evidence for broad Balmer features.\\

\noindent {\it eRASSt J045650.3{–}203750 \citep[][]{j0456_ptde}:} eRASSt J045650.3{–}203750 (J0456) was first discovered as a repeating, nuclear, X-ray transient by eROSITA. The X-ray lightcurve is characterized by a rise followed by a ${\sim}$months-long plateau that rapidly fades. This flare profile repeats every ${\sim}223$ days and has repeated three times. The host galaxy has no detectable optical broad or narrow spectral features, suggesting that it is quiescent in all senses. UV follow-up detected moderate variability, but no optical variability is detected. Radio emission was detected during the X-ray plateau. \\

\noindent {\it AT 2018fyk \citep[][]{fyk_discovery, fyk_ptde}:} AT 2018fyk was first detected as a nuclear optical transient by the ASASSN survey. Optical spectra taken post-flare detected broad He, He, and Bowen lines, along with narrow Fe\,II features. AT 2018fyk was detected as both a UV and X-ray transient in follow-up ($L_X \approx 10^{43}$\,erg\,s$^{-1}$, $L_{\rm UV, peak} = 10^{44}$\,erg\,s$^{-1}$) and the UV/X-ray emission remained bright for ${\sim}600$ days. After this period, the UV and X-ray emission plummeted, although remained detectable (plausibly due to a disk state change). ${\sim}1200$ days after the first flare, the UV and X-ray rebrightened, with $L_X \approx 8\times10^{42}$ erg s$^{-1}$ and $L_{\rm UV} \approx 10^{43}$\,erg\,s$^{-1}$. The host galaxy is classified as a retired galaxy based on its narrow optical emission lines. There is no evidence for AGN activity from infrared observations. It has a massive blackhole, with $M_{\rm BH} \approx 10^{7.7}\,M_\odot$.\\

\noindent {\it RX J133157.6–324319.7 \citep[][]{J1331_discovery, J1331_ptde}:} RX J133157.6–324319.7 (RX J1331) was detected as an ultra-soft X-ray transient with $L_X \approx 10^{43}$\,erg\,s$^{-1}$ by ROSAT in 1993 \citep[][]{J1331_discovery}. The transient evolved rapidly, with a factor of eight luminosity increase within ${\sim}8$ days. The host galaxy is non-active and non-star forming. It was not detected in sparse, survey observations until ${\sim}30$ years later, when it flared back to ${\sim}10^{43}$\,erg\,s$^{-1}$ during eROSITA observations. The X-ray spectrum during this flare was ultra-soft, like the first flare. It was not detected in follow-up observations ${\sim}20{-}100$ days post-peak. \\

\begin{deluxetable*}{c c c c c}
\centerwidetable
\tablewidth{10pt}
\tablecaption{Summary of published repeating pTDEs \label{tab:pTDEs}}
\tablehead{ \colhead{Name} & \colhead{Band} & \colhead{Period [years]} & \colhead{$\log M_{\rm BH}/M_\odot$} & \colhead{N. flares}}
\startdata
ASASSN-14ko$^{1,2,3}$ & Optical/UV/X-ray & 0.3 & 7.9 & ${\gtrsim} 20$ \cr
eRASSt J045650.3{–}203750$^{4}$ & X-ray/UV & 0.61 & 7.0 & 3 \cr
\hline
AT\,2020vdq$^{5}$ & Optical/UV & 2.7 & 6.1 & 2 \cr
\hline
AT 2018fyk$^{6,7}$ & UV/X-ray & 3.3 & 7.7 & 2 \cr
RX J133157.6–324319.7$^{8,9}$ & X-ray & ${\lesssim} 30$ & 6.5 & 2 \cr
\enddata
\tablecomments{Summary of published repeating pTDE candidates, sorted by flare period. The bands listed are those in which the rebrightening was detected. $^1$\cite{14ko_discovery}, $^2$\cite{14ko_xrayUV}, $^3$\citealp{14ko_followup}, $^{4}$\citealp{j0456_ptde}, $^5$This work, $^6$\citealp{fyk_discovery}, $^7$\citealp{fyk_ptde}, $^8$\citealp{J1331_discovery}, $^9$\citealp{J1331_ptde}}
\centering
\end{deluxetable*}

In Table~\ref{tab:pTDEs}, we tabulate the bands in which these TDEs were detected, their periods, and their black hole masses. There are a number of key facts to note about this sample:\\

\noindent \textit{Number of flares:} ASASSN-14ko is the only source that has been detected more than $2{-}3$ times; in fact, it has flared ${\sim}20$ times. It is the oldest source, which may partly explain the large number of detected flares. However, ZTF is sensitive to ASASSN-14ko-like objects out to $z\sim0.1$ and has been active long enough to detect ten flares from such an object, yet none have been reported. This may be a selection effect: the ZTF TDE selection requires that the transient not be present during the reference image, whereas an ASASSN-14ko-like source is on a larger fraction of the time than repeating pTDEs with longer periods. Moreover, the ZTF TDE selection will not include sources that repeated many times in the early months of the survey, so such events would not be included in the \cite{Yao_tdes} sample. Regardless, ASASSN-14ko is clearly distinct from the other pTDE candidates. The lack of detections of more such events may suggest that most pTDEs do not flare more than a few times. Alternatively, it may be a selection effect.\\

\noindent \textit{Black hole masses:} Of the optically-detected pTDEs (AT\,2020vdq, ASASSN-14ko, AT 2018fyk), all but AT\,2020vdq have black hole masses $\log M_{\rm BH}/M_\odot > 7$. The BH mass function of TDEs in ZTF begins to drop off around $\log M_{\rm BH}/M_\odot \sim 7.4$, and peaks towards masses lower than $\log M_{\rm BH}/M_\odot \sim 7$ \citep[][]{Yao_tdes}. In contrast, AT\,2020vdq has a black hole mass that is relatively typical for optically-selected TDEs. The host galaxies of ASASSN-14ko and AT 2018fyk are also very different from typical TDE host galaxies: ASASSN-14ko is a Seyfert, while AT 2018fyk is a retired galaxy with evidence for an old stellar population. In contrast, AT\,2020vdq shows strong Balmer absorption suggestive of ${\sim}$Gyr old stellar populations. This may suggest that types of stars (or binaries) being disrupted is different for ASASSN-14ko/AT 2018fyk and AT\,2020vdq.\\

\noindent \textit{Flare periods:} The flare period distribution (excluding RX J1331 due to a lack of constraints) appears to be bimodal, with one peak at ${\sim}100$s of days and one peak near ${\sim}2{-}3$ years. As has been detailed in other work \citep[e.g.][]{Cufari_hills}, placing stars on such short period orbits requires Hills capture of a binary. In this scenario, a binary is destroyed by the SMBH tidal forces, leading to one object being captured into a tight orbit around the SMBH and the other ejected. The typical binaries that will undergo this process will be on tight orbits themselves, as is required for them to survive the high velocity dispersion environment near the SMBH. These orbits are sufficiently tight as to require either two low mass stars or one star and one compact object. We will discuss the consequences of this fact further in Section~\ref{sec:rates}.\\

\noindent \textit{Multiwavelength properties:} The multiwavelength properties of the known pTDEs are very inhomogeneous. RX J1331 and J0456 are both primarily detected in the X-rays, with possible UV variability in the case of J0456. AT\,2020vdq, in contrast, shows little X-ray emission. Even when only comparing the optical/UV properties of ASASSN-14ko, AT\,2020vdq, and AT 2018fyk, we see a wide range of evolution and blackbody fit parameters, as is highlighted in Figure~\ref{fig:TRevol}. One possible common thread is that some, though not all, of the pTDE flares evolve more rapidly than typical TDEs.

\subsection{Constraints on TDE models from AT\,2020vdq}

Partial TDEs provide a unique opportunity to constrain both pTDE, but also fTDE models: the only aspects of the system we expect to change between disruptions are the stellar structure and, possibly, the CNM. With this in mind, we use the changes in the emission between flares from AT\,2020vdq to constrain both pTDE and fTDE models. There are four key takeaways from the observations of AT\,2020vdq:

{\noindent \textit{1. Launching a radio-emitting outflow may require a sub-Eddington state} \\}
\noindent AT\,2020vdq launched a radio outflow in the ${\sim}2$ years following its initial flare. Launching such an outflow requires an accretion disk. 
AT\,2020vdq thus likely had a pre-existing accretion disk during its rebrightening, and that this disk was, at one point, capable of launching radio-emitting outflows. Despite this, no new radio component was detected in the weeks following the rebrightening, when the event was in a relatively high accretion rate state (we are pursuing continued monitoring). There are three possible explanations for the lack of radio emission: (1) AT\,2020vdq-like TDEs cannot launch radio-emitting outflows except in a sub-Eddington state; (2) radio emitting TDEs can clear out the circumnuclear region in a region around the SMBH, so no material was present during AT\,2020vdq's rebrightening against which the outflow could shock; (3) the change in stellar structure affected the disk's ability to launch an outflow.  \\

{\noindent \textit{2. Bright and fast rebrightening \\}} 
\noindent The rebrightening of AT\,2020vdq was rapidly evolving relative to most TDEs. It has the fastest rise time and one of the fastest decay times of all TDEs. When controlling for rise time, it is also luminous. pTDE models predict that these events should evolve more rapidly than fTDEs: the fallback rate of stellar debris from fTDE decays is expected to follow a ${\sim}t^{-5/3}$ power-law, whereas pTDE decays are expected to follow a ${\sim}t^{-9/4}$ power-law due to the gravitational pull of the surviving stellar core. Intriguingly, however, the first flare from AT\,2020vdq was less extreme: the decay time was among the longest for the \cite{Yao_tdes} sample. This suggests that while pTDEs can be fast, they need not produce fast flares. It is possible that the fast evolution of the rebrightening is tied to the change in the stellar structure, rather than the pTDE nature of the event. However, \cite{Lodato_fallback} simulated the fallback rate for stellar disruptions with a range of polytropic indices and did not predict any early time fallback rates that were significantly steeper than the standard ${\sim}t^{-5/3}$. Instead, maybe the rapid evolution is due to a pre-existing accretion disk: the first pTDE episode likely created a disk with a size ${\sim}R_T$, and the interaction between a disk and stellar debris can cause rapid circularization and accretion \citep[][]{agn_tde}. \\

{\noindent \textit{3. Initial cooling of the optical flare\\}} 
\noindent The temperature evolution of AT\,2020vdq, while not unusual, deserves note. TDEs are often assumed to not cool. However, multiple TDEs show initial cooling near-peak, followed by ${\sim}10$s of days in a ${\sim}$constant temperature state. This has been noted by previous authors \citep[e.g.][]{qiz, 2020neh}, although models do not exist to explain the evolution to our knowledge. This evolution is particularly apparent from AT\,2020vdq: Near-peak, the temperature cools dramatically. It then hovers near a zero-cooling state.\\

{\noindent \textit{4. Otherwise, AT\,2020vdq is a fairly normal event\\}} 
\noindent The final point worth noting about AT\,2020vdq is that it is a relatively normal TDE. While it shows some unusual features (fast evolution, broad emission lines), it is not massively outside the typical TDE parameter space in any case. Both flares from this source would pass the ``normal'' TDE cuts used in \cite{Yao_tdes}. This suggests that some of the known TDEs could well be pTDEs.

\subsection{Partial TDE rates and the missing energy problem} \label{sec:rates}

We conclude by discussing the rate of partial TDEs. We then apply this discussion to the enigmatic missing energy problem of TDEs.

We first consider the rate of pTDEs, where each pTDE, regardless of the number of flares, is counted as a single event. Because AT\,2020vdq was drawn from the \cite{Yao_tdes} sample, we begin with the TDE rate measured in that work and scale it to estimate the pTDE rate. We first searched that sample for other pTDEs. One other transient, AT 2021mhg, showed a rebrightening. However, as we show in Appendix~\ref{app:mhg}, this source is a TDE followed by a Type Ia supernova, so we do not consider it further. AT\,2020vdq is the only definitive pTDE in the \cite{Yao_tdes} sample.

\cite{Yao_tdes} measured a TDE rate of ${\sim}3.2 \times 10^{-5}$ yr$^{-1}$ galaxy$^{-1}$ using $33$ ZTF TDEs. Note that this rate is an average over the population, but the rate depends on luminosity and increases for lower luminosities. Assuming that none of the other TDEs in this sample will rebrighten, the rate of partial TDEs on ${\sim}3$ year timescales is ${\approx}1/33 \times {\sim}3.2 \times 10^{-5}$ yr$^{-1}$ Galaxy$^{-1} \approx 10^{-6}$ yr$^{-1}$ galaxy$^{-1}$. This is a lower limit on the rate, as more rebrightening TDEs will increase the value, as will accounting for the luminosity dependence of the ZTF TDE rate and the ASASSN-14ko-like events that would not make it into the \cite{Yao_tdes} sample.

Instead of assuming that no more TDEs are hidden pTDEs, we now constrain the fraction of the \cite{Yao_tdes} that could be pTDEs. We assume that the pTDE flare period distribution is a uniform distribution ranging from $0.3{-}2.7$ years (i.e., all flares periods are between that of ASASSN-14ko and AT\,2020vdq). This flare period distribution was chosen arbitrarily and could affect our results: further theoretical studies are needed to motivate the correct form of this distribution. Assume the fraction of pTDEs in the \cite{Yao_tdes} sample is $f_{\rm pTDE}$. We assign each TDE in the \cite{Yao_tdes} sample to be a pTDE or fTDE according to this fraction. For those assigned as pTDEs, we randomly generate a flare period and count the number of pTDEs that would be observed to repeat before June 2023 (approximately when we last searched for ZTF pTDEs). We repeat this process $10^6$ times and calculate the fraction of trials where one or fewer objects are observed to repeat, which we call $P(N_{\rm pTDE}\le1)$. We find that $P(N_{\rm pTDE}\le1) < 10^{-3}$ for $f_{\rm pTDE}{\gtrsim} 0.3$; i.e., we have constrained the fraction of pTDEs in the \cite{Yao_tdes} to be ${\lesssim}30\%$ at the $3\sigma$ level. Thus, we can once more constrain the repeating pTDE rate for ${\sim}0.3-2.7$ year timescales to be ${\lesssim}30\%$ of the TDE rate, or ${\lesssim}10^{-5}$ yr$^{-1}$ galaxy$^{-1}$. Combining this result with the lower limit on the rate computed by assuming that AT\,2020vdq is the only pTDE in the \cite{Yao_tdes} sample, we find that the repeating pTDE rate for ${\sim}0.3-2.7$ year timescales is in the range $10^{-(5-6)}$ yr$^{-1}$ galaxy$^{-1}$.

Note that this result is strongly dependent on the assumed flare period distribution. Decreasing the lower limit on this distribution or having it peak towards smaller values tightens our constraints. However, if maximum allowed flare period is larger, a larger pTDE fraction is allowed. For example, if we assume the flare periods are drawn from a uniform distribution ranging from $0.3{-}5$ years, we can only say that is the pTDE fraction ${\lesssim}40\%$ at the $3\sigma$ level. One way to constrain the potential for longer timescale pTDEs is to look at older TDE samples, although no uniformly selected such sample exists. Regardless, we checked all the older optical/UV TDEs listed in \cite{suvi_review} for rebrightenings in ZTF or ATLAS data and do not find any detections. This suggests that TDEs do not frequently brighten on longer timescales, although this sample is small and inhomogeneous, so we do not attempt to set any detailed constraints on the longest allowed flare periods. We suggest, however, that at least some of the known TDEs are true, full TDEs rather than pTDEs.

In the theoretical frame work of Hills mechanism \citep{lu23_QPE_pTDEs}, the rate of repeating pTDEs depends on the rate at which stars migrate into deeply penetrating orbits and the binary fraction of stars in the galactic nucleus.
From the above discussion, we can conclude that the ratio of the number of Hills-captured, partially disrupted stars to tidally disrupted single stars is ${\sim}6\%$. This fraction is much less than unity mainly because few binaries can survive in galactic nuclei with semi-major axes that would produce a ${\sim}$year period pTDE.
Simulations of this process assuming different binary fractions and semi-major axis distributions would provide valuable information for interpreting the observed rate.

This measured rate, which suggests that pTDEs may not dominate the TDE rate, as well as the discussion in the previous parts of this work, have intriguing implications for the long-standing ``missing energy problem'' of TDEs: the total energy observed to be emitted during TDEs is always ${\sim} 2$ orders of magnitude lower than expected from the accretion of a ${\sim}$solar mass star. A number of possible explanations abound; for example, the bolometric corrections adopted to convert the observed luminosity into the total luminosity may be hugely overestimated \citep[e.g.][]{wenbin_missingE, mummery_model, mummery_missingE, saxton_xray, Guolo_xray}, more energy may be input into winds and/or jets than is currently expected \citep[e.g.][]{wenbin_missingE,dai_unified}, or the energy may be trapped in a disk with high optical depth \citep[][]{curd_grmhd_tde}. Alternatively, if all or most of the known TDEs are in fact pTDEs, the low detected energies would be expected because only a small fraction of the star need be accreted during each disruption. Our suggestion above that at least some of the known TDEs are true full TDEs refutes this hypothesis. Of course, if longer period (${\sim}$decade) pTDEs are common, it is still possible that pTDEs can aid the missing energy problems. Intriguingly, this pTDE hypothesis is supported by the fact that the original flare from AT\,2020vdq was originally identified as relatively typical TDEs, with no extremely unusual features that could have hinted at the fact that they would rebrighten, supports this possibility: the multiwavelength properties of the known pTDEs are so similar to the known TDEs that we can not rule out that any given event is instead a partial disruption. Future work on the long-term lightcurves of known TDEs will thus be critical for constraining the contribution of pTDEs to the missing energy problem. 

\section{Conclusions} \label{sec:conc}

In this work, we have presented observations and analysis of the partial tidal disruption event, AT\,2020vdq. AT\,2020vdq is the first pTDE identified from a well-characterized TDE sample, allowing us to constrain the pTDE rate to ${\approx}10^{-(5-6)}$ galaxy$^{-1}$ yr$^{-1}$, subject to uncertainties in the flare rate of pTDEs. We have used the properties of AT\,2020vdq to constrain TDE and pTDE models. Of note, the lack of a launch of a radio-emitting outflow shortly after the second disruption despite the launch of such an outflow after the first disruption may suggest that some TDEs must be in a low accretion state to launch such outflows. 

Intriguingly, and unlike previous pTDEs, the multiwavelength properties of AT\,2020vdq are generally characteristic of ``typical'' TDEs; for example, it has a low black hole mass and its optical flares are not hugely divergent from those observed from other TDEs (it was first identified as part of a standard TDE sample). This suggests that some of the known TDEs may in fact be pTDEs in disguise, so we urge careful monitoring of all events. However, we also urge caution when classifying repeating nuclear transients. As in the case of AT 2021mhg (Appendix~\ref{app:mhg}), large TDE samples combined with many-year monitoring campaigns leads to a high probability of chance detections of multiple, independent transients in a single host.

\begin{acknowledgments}

Based on observations obtained with the Samuel Oschin Telescope 48-inch and the 60-inch Telescope at the Palomar Observatory as part of the Zwicky Transient Facility project. ZTF is supported by the National Science Foundation under Grants No. AST-1440341 and AST-2034437 and a collaboration including current partners Caltech, IPAC, the Weizmann Institute of Science, the Oskar Klein Center at Stockholm University, the University of Maryland, Deutsches Elektronen-Synchrotron and Humboldt University, the TANGO Consortium of Taiwan, the University of Wisconsin at Milwaukee, Trinity College Dublin, Lawrence Livermore National Laboratories, IN2P3, University of Warwick, Ruhr University Bochum, Northwestern University and former partners the University of Washington, Los Alamos National Laboratories, and Lawrence Berkeley National Laboratories. Operations are conducted by COO, IPAC, and UW.

The ZTF forced-photometry service was funded under the Heising-Simons Foundation grant \#12540303 (PI: Graham).

Some of the data presented herein were obtained at the W. M. Keck Observatory, which is operated as a scientific partnership among the California Institute of Technology, the University of California and the National Aeronautics and Space Administration. The Observatory was made possible by the generous financial support of the W. M. Keck Foundation.

The authors wish to recognize and acknowledge the very significant cultural role and reverence that the summit of Maunakea has always had within the indigenous Hawaiian community.  We are most fortunate to have the opportunity to conduct observations from this mountain.

This material is based upon work supported by the National Science Foundation Graduate Research Fellowship under Grant No. DGE‐1745301. MWC acknowledge support from the National Science Foundation with grant numbers PHY-2010970 and OAC-2117997. MN is supported by the European Research Council (ERC) under the European Union’s Horizon 2020 research and innovation programme (grant agreement No.~948381) and by UK Space Agency Grant No.~ST/Y000692/1.

This research is based on observations made with the NASA/ESA Hubble Space Telescope obtained from the Space Telescope Science Institute, which is operated by the Association of Universities for Research in Astronomy, Inc., under NASA contract NAS 5–26555. These observations are associated with program(s) 17314.
\end{acknowledgments}
\clearpage

\appendix

\section{The second repeating ZTF transient, AT 2021mhg: a TDE and SN Ia in the same galaxy} \label{app:mhg}

\begin{figure}[b!]
    \centering
    \includegraphics[width=\textwidth]{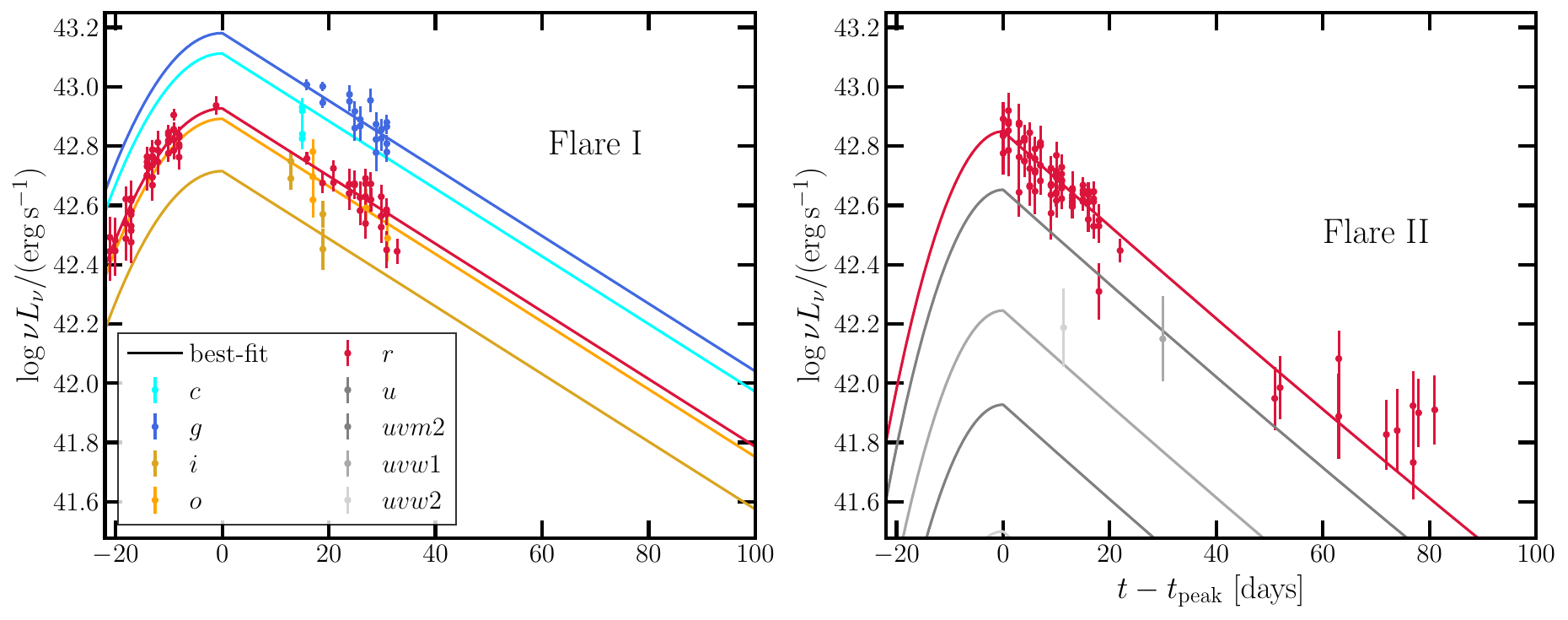}
    \caption{Optical/UV lightcurve of AT 2021mhg, in the same format as Figure~\ref{fig:opt_lc}.}
    \label{fig:at2021mhg_lc}
\end{figure}

After the discovery of AT\,2020vdq, we performed forced photometry for all ZTF TDEs to search for any other rebrightenings on ${\sim}$year timescales. Coincidentally, AT 2021mhg had rebrightened in the $r$-band shortly before AT\,2020vdq \citep[][]{mhg_tns}. Upon discovery, we and other groups triggered multiwavelength follow-up. In this section, we present observations of this source and constrain its origin. We conclude that this source is a TDE followed by a Type Ia supernova at the same location.

\subsection{Host galaxy}

AT 2021mhg is hosted by a Balmer-strong, blue galaxy with a black hole mass $\log M_{\rm BH}/M_\odot = 6.1 \pm 0.37$. No narrow emission lines are detected in the host spectra, although no host galaxy spectra without any transient emission are available. Strong host absorption lines are detected, in contrast. We tentatively identify this galaxy as quiescent, both in terms of star formation and AGN activity.

\begin{deluxetable*}{c|cccccc}[b!]
\centerwidetable
\tablewidth{10pt}
\tablecaption{ \label{tab:optlc_mhg}}
\tablehead{ \colhead{} & \colhead{$t_{\rm peak}$} & \colhead{$\log \frac{L_{\rm bb}}{{\rm erg\,s}^{-1}}$} & \colhead{$\lambda_{\rm Edd.}$} & \colhead{$\log T_{\rm bb}/{\rm K}$} & \colhead{$t_{\rm 1/2,rise}$} & \colhead{$t_{\rm 1/2,decay}$} \cr 
 \colhead{} & \colhead{} & \colhead{} & \colhead{} & \colhead{} & \colhead{[day]} & \colhead{[day]}}
\startdata
Flare I & $59366.6 \pm 0.7$ & $43.65 \pm 0.05$ & $0.28 \pm 0.03$ & $4.21 \pm 0.03$ & $9.7 \pm 0.4$ & $26 \pm 2$\cr
Flare II & $60072.5$ (fixed) & $43.00 \pm 0.02$ & $0.063 \pm 0.002$ & $3.82 \pm 0.02$ & $10$ (fixed) & $18.8 \pm 1.1$
\enddata
\tablecomments{Best-fit evolving blackbody parameters for the optical flares from AT 2021mhg.}
\centering
\end{deluxetable*}
\clearpage
\subsection{Optical/UV broadband emission}

The optical/UV emission associated with the initial and secondary flares are shown in Figure~\ref{fig:at2021mhg_lc}. The secondary flare was only observed with ZTF/$r$, but we have additional Swift/UVOT photometry (PIs Somalwar, Huang). In strong contrast with the rebrightening of AT\,2020vdq, the second flare from AT 2021mhg is significantly redder than the first flare. The rebrightening also has a slightly lower luminosity. Like AT\,2020vdq, the second flare fades faster than does the first, but the contrast is not as extreme as that of AT\,2020vdq.

We model the optical/UV flares following the same methods as used for AT\,2020vdq, with small modifications. First, we find that no temperature evolution is required for either flare, so we fix the time derivative of the temperature to zero. Second, the rise of the rebrightening was not observed by any telescopes. Hence, we fix the peak MJD to the date of the first detection and we arbitrarily fix the rise time to $10$ days (this choice of rise time does not affect our results). The resulting best-fit lightcurve parameters are tabulated in Table~\ref{tab:optlc_mhg}, and the best-fit models are shown in Figure~\ref{fig:at2021mhg_lc}. As expected, the second flare is $0.4$ dex cooler than the first flare; in fact, AT 2021mhg's rebrightening would be the reddest known optical TDE flare. It also has a very low luminosity, with an Eddington ratio $<0.1$. 

\begin{figure*}
    \centering
    \includegraphics[width=\textwidth]{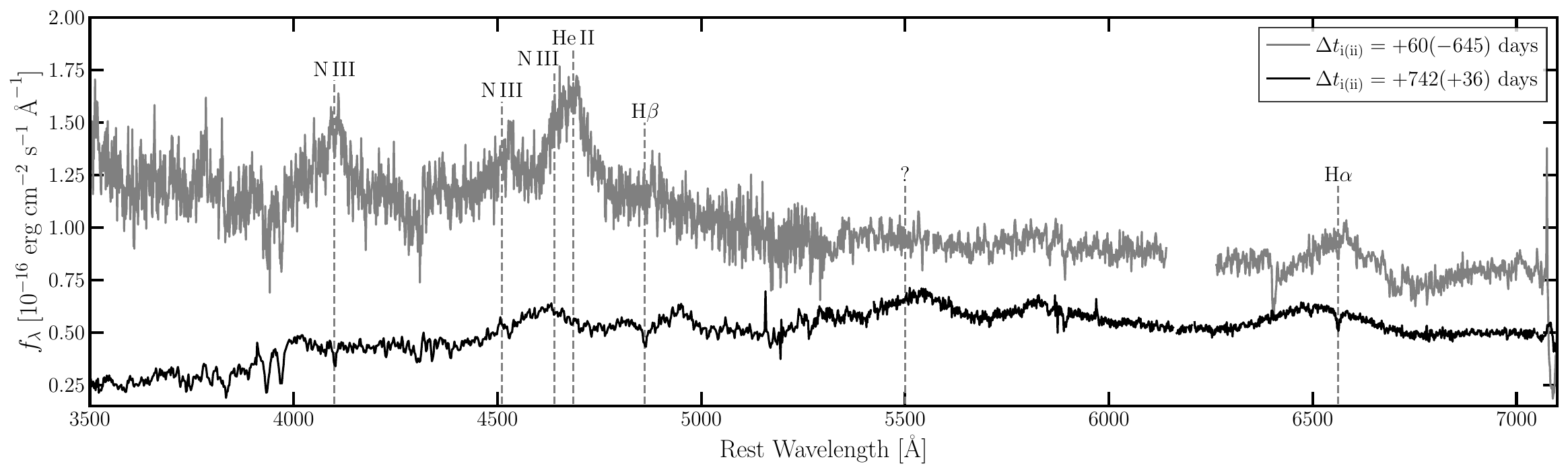}
    \caption{Spectral evolution of AT 2021mhg. The grey spectrum was observed $60$ days after the first flare and the black spectrum was observed $36$ days after the second flare. }
    \label{fig:at2021mhg_spec}
\end{figure*}



We also performed a non-parameteric fit to those epochs with UV data and confirm the result that the rebrightening was extremely cool: no previously known TDE with non-parametric modelling was cooler. 

\subsection{Optical Spectral Features}

We next consider the optical spectra of AT\,2020vdq. In Figure~\ref{fig:at2021mhg_spec}, we show one spectrum of AT\,2020vdq observed with P200/DBSP (PI: Kulkarni; \citep[][]{Yao_tdes}) ${\sim}70$ days after the first flare (grey) and one spectrum observed ${\sim}35$ days after the rebrightening with Keck I/LRIS (black). These spectra look markedly different. Broad Balmer, He\,II, and N\,III are detected after the first flare. Note that the line near $4100\,{\rm \AA}$ is assumed to be N\,III$\lambda 4100$ rather than H$\delta$; it would be theoretically difficult to produce such bright H$\delta$ relative to H$\beta$ and H$\alpha$, but the detection of N\,III is fully consistent with TDE models. The detection of these lines places AT 2021mhg in the H$+$He TDE class. The presence of N\,III places AT 2021mhg in the class of Bowen TDEs, which show Bowen fluorescence lines. The detected lines do not show significant velocity offsets and have widths ${\sim}8000$ km s$^{-1} = 0.03c$. These line widths are typical for H$+$He TDEs that are observed ${\sim}50$ days post-peak.

The post-rebrightening spectrum is extremely unusual for a TDE. We detect the typical broad lines He\,II$\lambda 4686$, possible H$\beta$ and He\,I$\lambda 5876$, and H$\alpha$. However, there are also broad features near $5000\,\AA$ and $5500\,\AA$ that, to our knowledge, have not been seen from a TDE before and that we cannot identify. N\,III$\lambda 4100$ is not obviously present. The detected lines are broader than those in the previous spectrum, with FWHM$\approx11000$ km s$^{-1} \approx 0.04c$. This line width is comparable to that measured for other of non-Bowen TDEs (i.e., those with no detectable N\,III lines). They are also significantly blueshifted by $\Delta v \approx 3000$ km s$^{-1}$.

\subsection{Radio and X-ray limits}

\begin{deluxetable}{cc}
\centerwidetable
\tablewidth{10pt}
\tablecaption{ Swift/XRT $5\sigma$ upper limits \label{tab:swift_mhg}}
\tablehead{ \colhead{MJD} & \colhead{AT 2021mhg} \cr
\colhead{} & \colhead{[$10^{40}$ erg s$^{-1}$]}}
\startdata
59417.7 & ${<}100.4$ \cr
59490.2 & ${<}3.2$ \cr
59501.1 & ${<}4.9$ \cr
59572.3 & ${<}58.7$ \cr
60083.8 & ${<}3.9$ \cr
60088.2 & ${<}3.8$ \cr
60095.1 & ${<}37.8$ \cr
60102.1 & ${<}3.7$ \cr
\enddata
\centering
\end{deluxetable}

Neither radio nor X-ray detections have been reported for either brightening of AT 2021mhg. X-ray upper limits from the Swift X-ray telescope are reported in Table~\ref{tab:swift_mhg}. 

AT 2021mhg was observed at 15 GHz $70$ days after the initial flare and an upper limit of $3\times 10^{37}$\,erg\,s$^{-1}$ was reported \cite{mhg_radio_TNS}. It was observed by the VLA Sky Survey at 3 GHz $130$ days after the initial flare and an upper limit of $3\times 10^{38}$\,erg\,s$^{-1}$ was reported. No post-rebrightening radio observations have been reported; we are pursuing follow-up.

\subsection{The cause of AT 2021mhg}

While the first flare from AT 2021mhg is a typical TDE, the second flare is highly unusual. In particular, it has a remarkably red color corresponding to a blackbody temperature that would be significantly cooler than all known optical TDEs. We also see broad features in the post-rebrightening optical spectrum that have never been detected from a TDE. Due to these features, we consider non-TDE origins for the rebrightening of AT 2021mhg. We do not see any evidence for AGN activity from AT 2021mhg (e.g., no narrow lines are detected in the optical spectrum); hence, if this source is not a pTDE, it is most likely a supernova. We use the Supernova Identification (\texttt{snid}) tool to model the post-rebrightening optical spectrum from this source. The spectrum is well-modeled ($87\%$ match) as a type Ia supernova observed ${\sim}30$ days post-peak. The mysterious broad features near $5000\,{\rm \AA}$ are perfectly modeled in this scenario. If this second flare is indeed a Type Ia SN, the red, fast-evolving lightcurve is expected.

Thus, the second flare from AT 2021mhg would be extremely unusual if it were caused by a TDE, but it is fully consistent with originating from a Type Ia SN. In this case, we have observed a galaxy where a TDE occured, followed by a Type Ia SN after ${\sim}2$ years. The probability of observing a scenario like this is non-neglible, and must be considered in future searches for pTDEs. A Type Ia SN in a ${\sim}$Milky Way-like galaxy once every ${\sim}500$ years. Assuming this rate is uniform throughout the galaxy, the typical TDE host has a radius of ${\sim}3\arcsec$and we define the nucleus as the central ${\sim}1\arcsec$, a Type Ia SN will occur in a galaxy nucleus every $2\times 10^4$ years. 

We have searched a sample of $33$ galaxies (i.e., the \citealp{Yao_tdes} sample) for flares over a period of ${\sim}3$ years (i.e., the time since the first TDE flares in each). Thus, the probability of detecting a Type Ia SN from one of these galaxies in this time period is $33\times 3\,{\rm years}/2\times 10^4\,{\rm years} = 0.02$. The probability of detecting one or more events given that this expected probability is ${\sim}2\%$. In other words, the detection of AT 2021mhg source is only a ${\sim}2\sigma$ fluctuation.

As we continue monitoring TDEs for rebrightenings, the probability of detecting both a TDE and another transient in the same galaxy will continue increasing. We strongly urge the careful consideration of this scenario when future pTDE candidates are detected.

\bibliography{sample631}{}
\bibliographystyle{aasjournal}
\end{document}